\documentclass[useAMS,usegraphicx]{mn2e}
\usepackage{rotate}
\usepackage{url}
\usepackage{rotating}
\usepackage{times}
\usepackage{cite}             
\usepackage{comment}  
\usepackage{graphicx}    
\usepackage{color}          
\usepackage{arydshln}    
\newif\ifAMStwofonts
\AMStwofontstrue

\def\mrk493{{Mrk~493}}
\def\1H{{1H\,0707$-$495}}
\def\iras{{IRAS~13224$-$3809}}

\def\kdblur2{\sc{kdblur2}}

\def\kdblur{\sc{kdblur2}}

\def\et{{et al.\ }}
\def\Rg{{\thinspace $R_{\rm g}$}}          

\def\Fvar{{$F_{\rm{var}}$}}

\def\chisq{{$\chi^2$}}               

\def\redchi{{$\chi^2_\nu$}}

\def\swift{{\it Swift }}
\def\xmm{{\it XMM-Newton }}

\def\feka{{Fe~K$\alpha$}}

\def\cm{{\rm\thinspace cm}}
\def\erg{{\rm\thinspace erg}}
\def\eV{{\rm\thinspace eV}}

\def\ga{{\rm\thinspace gauss}}

\def\keV{{\rm\thinspace keV}}
\def\km{{\rm\thinspace km}}

\def\Mpc{{\rm\thinspace Mpc}}

\def\s{{\rm\thinspace s}}
\def\arcsec{{\rm\thinspace arcsec}}   
\def\ks{{\rm\thinspace ks}}
\def\ps{{\rm\thinspace s^{-1}}}

\def\cts{{\rm\thinspace count}}

\def\cps{\hbox{$\cts\s^{-1}\,$}}

\def\ergcmps{\hbox{$\erg\cm\s^{-1}\,$}}                        
\def\ergpscmps{\hbox{$\erg\cm^{-2}\s^{-1}\,$}}
\def\ergps{\hbox{$\erg\s^{-1}\,$}}

\def\phpcmsps{\hbox{$\rm{ph}\cm^{-2}\s^{-1}\,$}}               

\def\kmps{\hbox{$\km\ps\,$}}
\def\kmpspMpc{\hbox{$\kmps\Mpc^{-1}\,$}}

\title[Variable blurred reflection in \mrk493]{Variable blurred reflection in the narrow-line Seyfert~1 galaxy \mrk493}

\author[K. Bonson et al.]{K. Bonson,$^1$ L. C. Gallo,$^1$ D. R. Wilkins,$^2$ A. C. Fabian,$^3$  
               \\ 
$^{1}$ Department of Astronomy and Physics, Saint Mary's University, 923 Robie Street, Halifax, NS, B3H 3C3, Canada \\
$^{2}$ Kavli Institute for Particle Astrophysics and Cosmology, Stanford University, Stanford, CA, 94305, U.S.A. \\
$^{3}$ Institute of Astronomy, University of Cambridge, Madingley Road, Cambridge, CB3 0HA, United Kingdom}
\date{Accepted. Received. }
\pagerange{\pageref{firstpage}--\pageref{lastpage}}
\pubyear{2016}
\begin{document}
\maketitle
\label{firstpage}

\begin{abstract}
{We examine a 200\ks\ \emph{XMM-Newton} observation of the narrow-line Seyfert~1 galaxy \mrk493. The active galaxy was half as bright as in a previous 2003 snapshot observation and the current lower flux enables a study of the putative reflection component in detail. We determine the characteristics of the 2015 X-ray continuum by first analyzing the short-term variability using model-independent techniques. We then continue with  a time-resolve analysis including spectral fitting and modelling the fractional variability. We determine that the variability arises from changes in the amount of primary flux striking the accretion disk, which induces changes in the ionization parameter and flux of the blurred reflection component.  The observations seem consistent with the picture that the primary source is of roughly constant brightness and that variations arise from changes in the degree of light bending happening in the vicinity of the supermassive black hole.
}
\end{abstract}


\begin{figure*}
   \centering
   \advance\leftskip-0.5cm
   {\scalebox{0.36}{\includegraphics[trim= 1.85cm 1cm 1.5cm 0cm, clip=true]{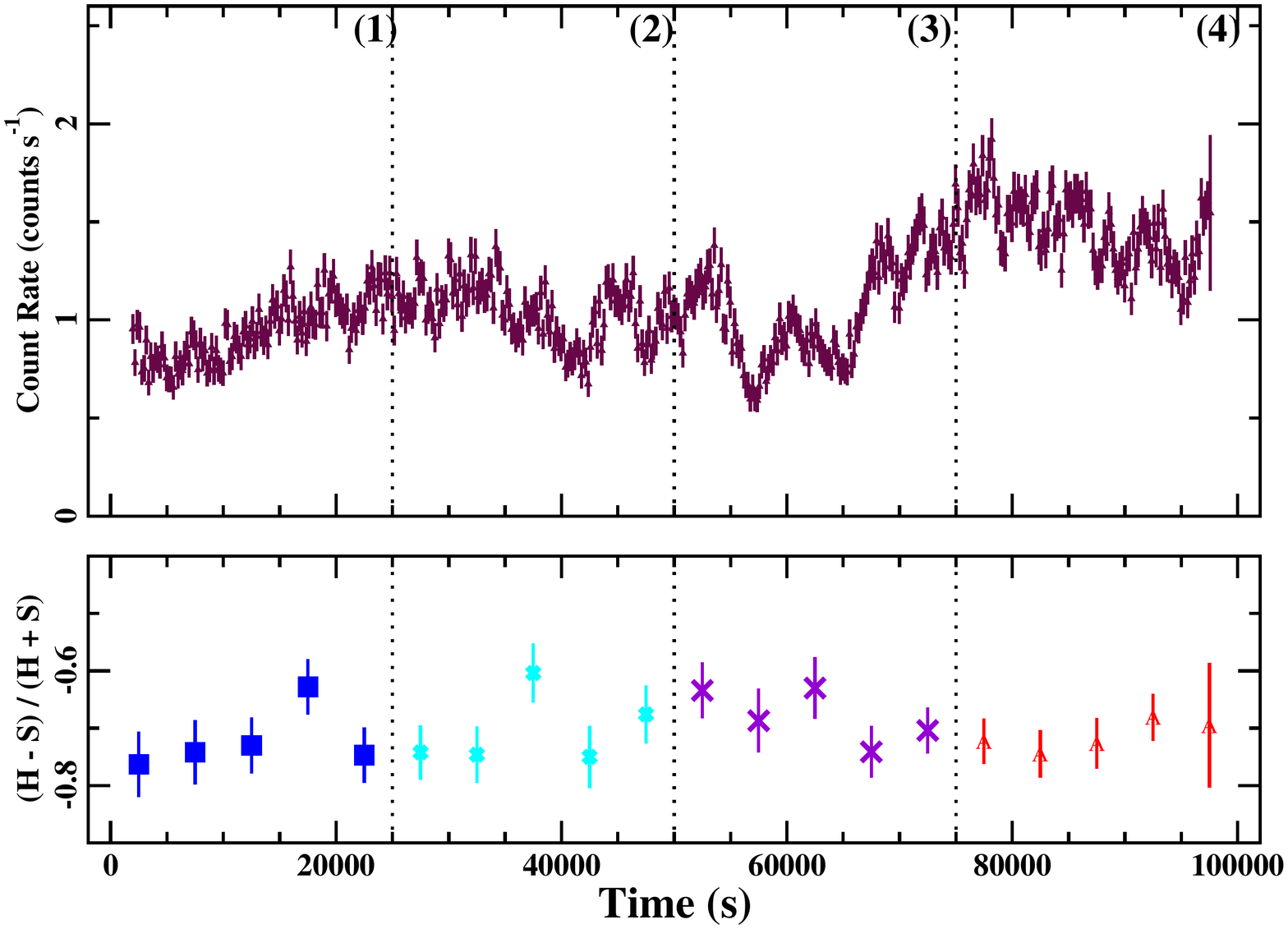}}}      
   {\scalebox{0.36}{\includegraphics[trim= 1.85cm 1cm 1.5cm 0cm, clip=true]{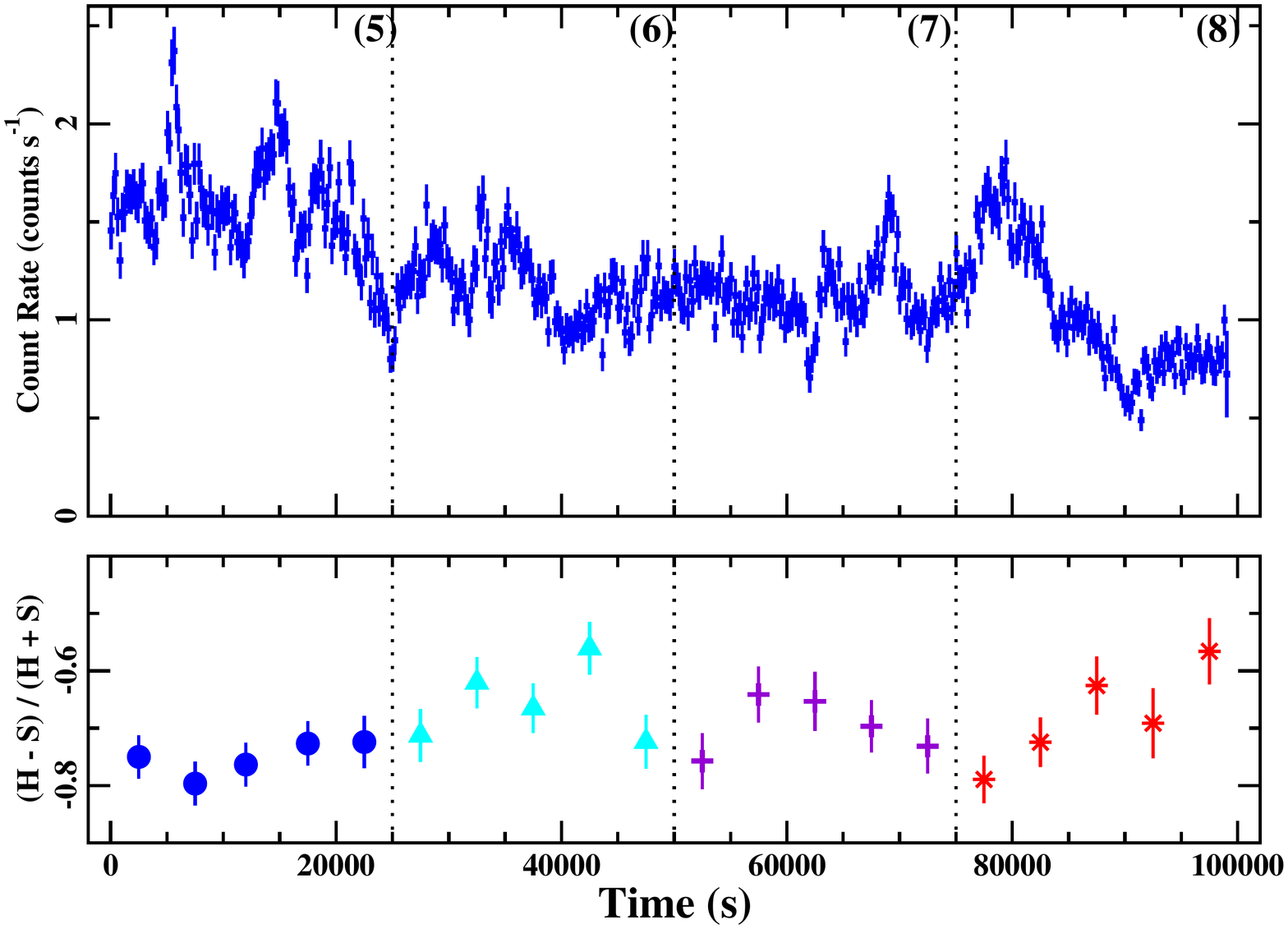}}}       
   \caption{The 0.2-8.0\keV\ merged-MOS light curves for XMM15a (top left; maroon triangles) and XMM15b (top right; blue squares).  The observations are separated by about $5$ days.
   The time segments adopted in the time-resolved analyses in Section~\ref{Ref} are shown. The XMM15 spectra were extracted every 25\ks\ for a total of four time bins per epoch. The divisions also serve as a rough flux-resolved analysis. 
  Time-resolved hardness ratios compare the soft (0.35-0.5\keV) and hard (5-8\keV) bands (lower panels).  The hardness ratios were derived with light curves binned by 5\ks.  The symbols in Bin 1 (squares), Bin 2 (hourglasses), Bin 3 (crosses), Bin 4 (three-pointed stars), Bin 5 (circles), Bin 6 (triangles), Bin 7 (pluses), and Bin 8 (stars) will be used in the time-resolved analysis.}
   \label{TRL}
\end{figure*}

\section{Introduction}
\label{intro}
Narrow-line Seyfert 1 (NLS1) galaxies are a class of active galactic nuclei (AGN) that are known for being exceptional at all wavelengths.  On average, they have higher star formation rates (e.g. Sani \et 2010) than the broad line Seyfert 1 galaxies. Some objects have even been identified with strong jet and $\gamma$-ray emission (e.g. Foschini \et 2012, 2015).  In X-rays, many objects possess spectra that are dominated by General relativistic effects (e.g. Fabian \et 2009, 2013; Ponti \et 2010) and exhibit extreme variability on various timescales (e.g. Grupe \et 2007, 2012; Gallo \et 2004).  It is generally believed NLS1s possess smaller supermassive black holes than normal Seyfert 1s and are accreting at high Eddington rates. 

Below about $2\keV$, the X-ray spectra of Type 1 AGN show a soft-excess above the power law continuum, which is often enhanced in NLS1s.  The nature of the soft excess is often debated (e.g. Ross \& Fabian 2005, Crummy \et 2006, Done et al. 2007, Done et al. 2012, Boissay et al. 2016) and one possibility is that the emission arises from blurred reflection of the primary source off the inner accretion disk (e.g. Ross \& Fabian 2005).  Measurements of reverberation lags lend support to this interpretation (e.g. Zoghbi \et 2010).  

The soft excess contains information on composition and the ionization of the inner disk.  Modelling the soft excess and observing how the component various in conjunction to the primary source reveals information on the nature of the disk (e.g. densities)  and the geometry (sizes) of the inner environment.  The strong soft excess associated with NLS1s makes it possible to accurately study the inner disc properties and behaviour.

At low X-ray flux levels, the opportunity to study the nature of the accretion disk is improved as the continuum flux from the power law component is diminished and the reflection spectrum is enhanced (e.g. Miniutti \& Fabian 2004).  NLS1s are ideal to study AGN at low X-ray flux levels (specifically X-ray weak states) because they show strong variations from their nominal X-ray brightness.  When they are X-ray bright, NLS1 tend to display simple, power law dominated spectra compared to when they are X-ray weak and exhibit more complex behaviour (e.g. Gallo 2006).

The NLS1 \mrk493\ is a local (z = 0.031) and bright AGN that is well studied at various wavelengths.  However, detailed studies of the NLS1 in the X-rays are few.  \mrk493\  has been viewed by \swift\ every few years since 2005 (e.g. Grupe \et 2010).  On average, the source is about twice as bright as \1H and \iras, but only various by a factor of about three over yearly timescales.  There has been no detection of \mrk493\ in the 105-month BAT survey (Oh \et 2018).   A pointed 13.7\ks\ \xmm observation in 2003 revealed an X-ray spectrum with a prominent soft-excess and  \feka\ emission line.  The source was categorized as a simple source by Gallo (2006) in that it did not exhibit significant spectral complexity.  

A deep $190\ks$ observation in 2015 with \xmm\ caught the source in a low X-ray flux state, which we report here. 
This work is organized as follows: Section \ref{data} summarizes the data collection and reduction. Section \ref{MIA} introduces the model-independent techniques first utilized to characterize the behaviour of \mrk493 in the X-ray.  A preliminary examination of the X-ray spectrum is completed in Section~\ref{meanspec}.   The blurred reflection model is tested on the X-ray spectrum and used to describe the variability in Section \ref{Ref}. Discussion and conclusions follow in Section \ref{discuss} and Section \ref{conclude}, respectively.


\section{Observations and data reduction}
\label{data}
The 2003 \xmm observation (ObsID 0112600801; hereafter XMM03) was during revolution 0568 starting on 16 January 2003 and spanned 19\ks. The pn camera (Str\"uder \et 2001) operated in large window mode and the two MOS detectors (Turner \et 2001) operated in small window mode, all three detectors used a medium filter.

The first 2015 \xmm observation (ObsID 0744290201; hereafter XMM15a) was during revolution 2786 starting on 24 February 2015 and spanned 97\ks. The second observation (ObsID 0744290101; hereafter XMM15b) was during revolution 2789 starting on 2 March 2015 and spanned 100.4\ks. The EPIC detectors all operated in small window mode with a medium filter for both 2015 observations. A log of the \xmm\ observations is presented in Table~\ref{ObsDetails}. 
\begin{table}
\caption[Mrk 493 Data Log]{\mrk493 \xmm data log for the single 2003 and two 2015 observations. Some background flaring was evident in XMM15b.}
\centering    
\scalebox{1.0}{                                      
		\begin{tabular}{ccccc}
		\hline
		(1) & (2) & (3) & (4) & (5)  \\
		ObsID & Start Date & Duration  & Instrument  & GTI  \\
		(Designation) &  &(ks) &   &  \rm{(ks)} \\\hline
		 0112600801 & 16/01/2003 & 19  & pn   & 13 \\
		 (XMM03) &&   & MOS 1 & 18 \\
		&&   & MOS 2 &18 \\
		 \\
		 0744290201 & 24/02/2015 & 97 & pn    & 66 \\
		  (XMM15a) && & MOS 1 &92\\
		&& & MOS 2 &92 \\
		\\
		 0744290101 &  02/03/2015 & 100 & pn  & 69  \\
		 (XMM15b) && & MOS 1 &87  \\
		&& & MOS 2 & 86  \\
		\\
		\hline
		\end{tabular}}
	\label{ObsDetails}
\end{table}

The Optical Monitor (OM; Mason \et 2001) and Reflection Grating Spectrometers (RGS1 and RGS2; den Herder \et 2001) observed simultaneously with the EPIC instruments during the 2015 epochs.  The RGS were of low signal-to-noise and did not exhibit significant features. These data will not be discussed further.  The OM data contain significant host-galaxy contribution and will be discussed elsewhere. 

Data files from all epochs were processed to produce calibrated event lists using the \xmm Science Analysis System (SAS) version 15.0.0 and latest calibration files at the time of writing. The data were examined for background flaring and pileup. Moderate flaring was seen throughout XMM15b and those time periods were ignored. No pileup was detected in either of the 2015 observations. Pileup was present in the 2003 pn data and this was corrected for.  Source photons were extracted from a circular region 35\arcsec\ in radius and centred on the object. The background photons were extracted from an area 50\arcsec\ in radius close to the object and then scaled appropriately.
 
Single to quadruple events were selected for the MOS data while single and double events were selected for the pn. Events next to a bad pixel or the CCD edge were omitted (i.e. data quality flag set to zero). The MOS spectra from the 2015 data were limited to the 0.3 -- 8.0\keV\ range because of high background at $E >$ 8\keV. The pn spectra of XMM15a were source dominated only  between 0.3 -- 6.5\keV, due to high background. The resulting mean count rates in the $0.3-10\keV$ band for XMM03 were: 5.86\cps\ (pn), 1.41\cps\ (MOS 1), and 1.43\cps\ (MOS 2).  For XMM15a  mean count rates were: 1.88\cps\  (pn), 0.45\cps\  (MOS 1), and 0.45\cps\  (MOS 2) in the 0.3 -- 8.0\keV\ band. For XMM15b the mean count rates were: 2.03\cps\  (pn), 0.50\cps\  (MOS 1), and 0.49\cps\  (MOS 2) in the same energy band as XMM15a.

In 2015, the MOS instruments provided good data over a wider bandpass than the pn because of higher background levels.  The pn data were analysed throughout this work to examine for consistency between the instruments, but only the MOS data are reported.

\begin{figure}
   \centering
   {\scalebox{0.3}{\includegraphics{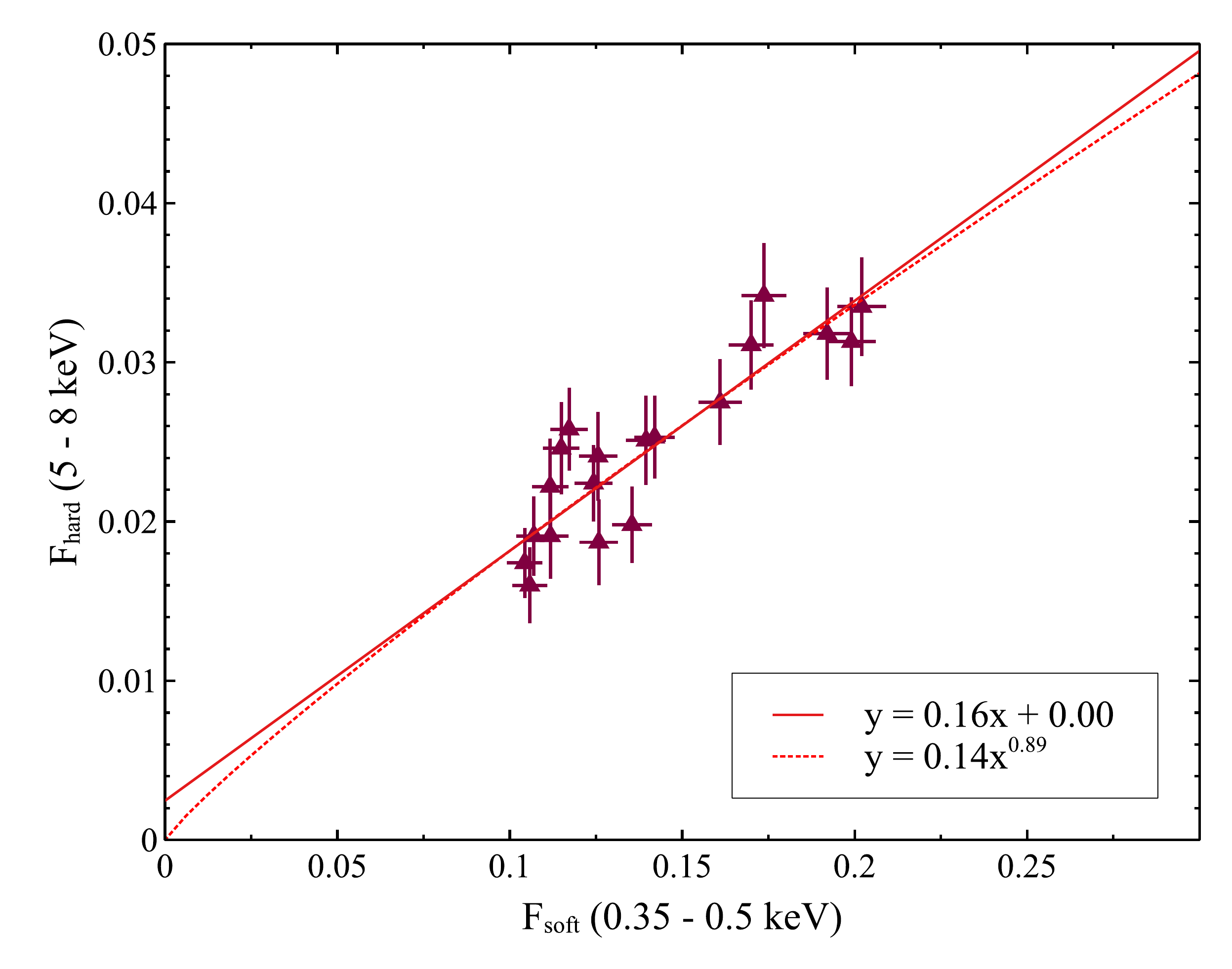}}}
   {\scalebox{0.3}{\includegraphics{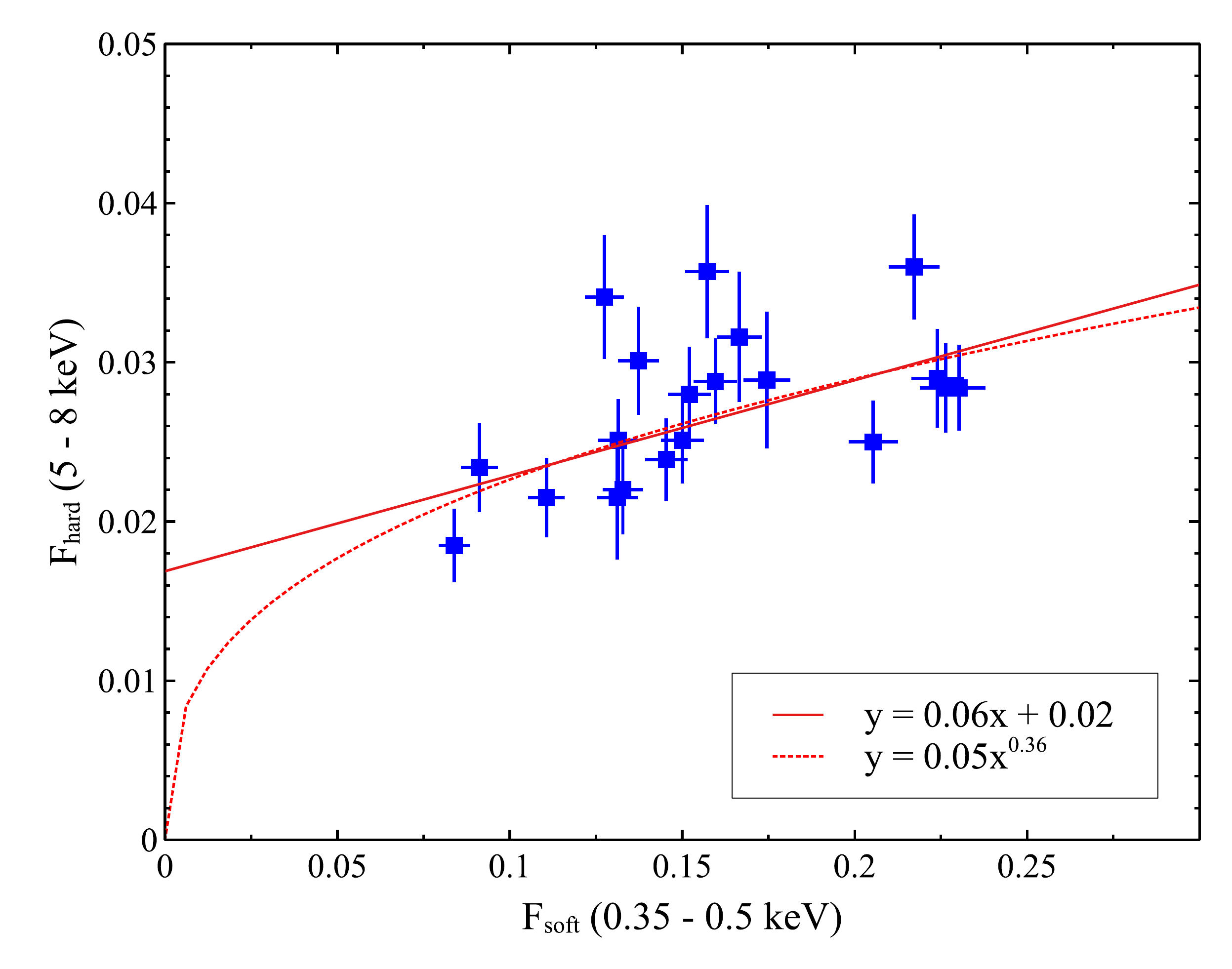}}}
   \caption{Flux-flux plots for XMM15a (top panel, maroon triangles) and XMM15b (lower panel, blue squares).  Both curves were fit with linear (solid red line) and power law (dotted red line) models. A linear fit is sufficient to describe XMM15a (top panel), whereas neither model describes XMM15b well.}
   \label{FvF}
\end{figure}


\section{Model-Independent Analysis}
\label{MIA} 
Before delving into detailed multi-epoch spectroscopy, the 2015 \mrk493 data are examined in a model-independent fashion. It is very common for several physical scenarios to describe spectra equally well (e.g. Bonson et al. 2015, Gallo et al. 2013, 2015) and so we hope to gather preliminary clues from the X-ray variability in \mrk493 that may aid in later spectral analysis.

Source photons were extracted to create light curves in various energy bands between 0.2 -- 10\keV\ for XMM03 and  0.2 -- 8\keV\ for XMM15. At each epoch, the light curves from all EPIC instruments agree, therefore the MOS 1 and 2 light curves were combined, taking care to match the start and stop times. 

The 2015, $0.2-8\keV$ merged-MOS light curves are shown in Fig. \ref{TRL}. A simple visual inspection of the 2015 light curves show that the source varies by about 50 per cent and on timescales as short as  $\sim$\thinspace5\ks. Similar behaviour is found in 2003, though the duration of the observation was much shorter than in 2015. 

Hardness ratios between the soft (0.35 -- 0.5\keV) and hard (5 -- 8\keV) bands in each observation were plotted with respect to time and reveal evidence of spectral variability  (Fig. \ref{TRL}). No significant correlation is found between hardness ratio and count rate.

\subsection{Flux-flux plots}

To  investigate the spectral variability, flux-flux plots comparing the $0.35-0.5\keV$ and $5-8\keV$ bands were created following Taylor et al. (2003) (Fig. \ref{FvF}).  The light curves were binned by 5\ks. 

The XMM15a flux-flux plot could be well fit with a linear model and a positive y-intercept (\redchi = 0.93).  This behaviour is consistent with a two-component continuum model where changes in the flux of a soft component (with constant shape) occur over a relatively constant hard component.  In the case of MCG--6-30-15, the hard constant component was interpreted as blurred reflection (e.g. Taylor \et 2003).  

On the other hand, the XMM15b flux-flux plot is fit poorly with both a linear (\redchi = 1.59) and power-law  (\redchi = 1.50) model.  This suggests the variability is more complex, perhaps requiring multiple components to vary in shape and flux at the same time. The transition from the simple behaviour in XMM15a and the more complicated behaviour in XMM15b  occurs in less than $\sim5$ days (i.e. the time between epochs). 

\begin{figure}
   \centering
   \advance\leftskip-0.3cm            
   {\scalebox{0.33}{\includegraphics[trim= 0cm 0cm 0cm 0cm, clip=true]{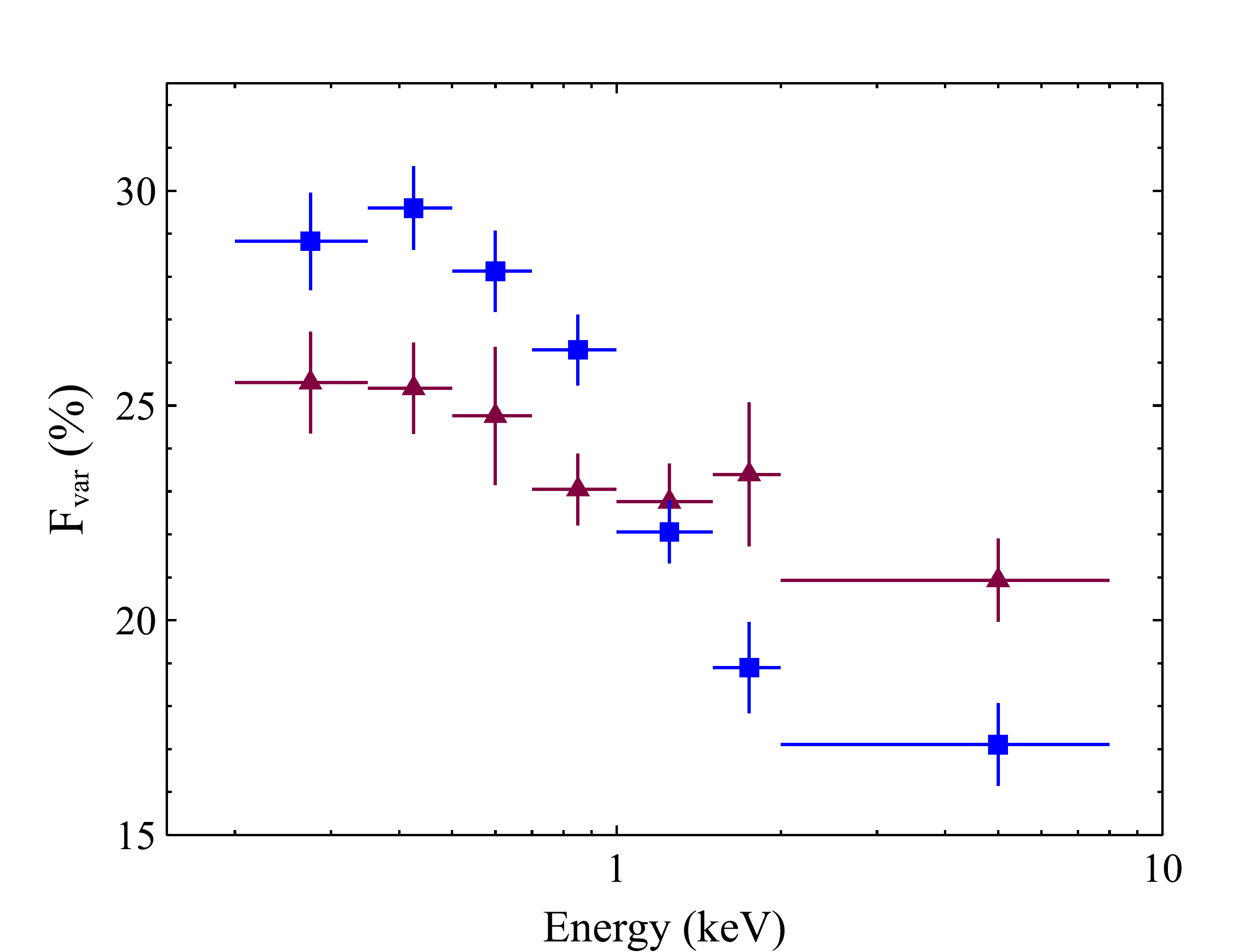}}}
   \caption{Fractional variability (\Fvar) of the merged-MOS broad band light curves. Both epochs show increased spectral variability in the soft ($<$1\keV) band, with the variability in XMM15b being far more significant. As in Fig. \ref{TRL}, maroon triangles and blue squares represent XMM15a and XMM15b, respectively.}
   \label{MOSFvar}
\end{figure}


\subsection{Fractional Variability}
\label{Av_Fvar}
Root-mean-squared fractional variability (\Fvar) analysis is used to quantify the variability intrinsic to a light curve while accounting for uncertainty (Edelson et al. 2002, Ponti et al. 2004). Mathematically,
%
\begin{equation}
F_{\rm{var}} = \frac{1}{\langle X \rangle}\sqrt{S^2 - \langle \sigma^2_{\rm{err}} \rangle}
\label{FvarEq}
\end{equation}
%
where $\langle X \rangle$ is the mean count rate, the total variance of the light curve is $S^2$, and $\langle \sigma^2_{\rm{err}} \rangle$ is the mean error squared. Calculating the \Fvar\ as a function of energy portrays the amplitude of the variations in different energy bands, therefore revealing spectral variability.  The uncertainties are determined following Ponti et al. (2004). 

The 2015 light curves were binned by $200\s$ and a total of seven energy bins between $0.2-8.0\keV$ were used to calculate the  \Fvar\ spectrum. 
The \Fvar\ spectrum in XMM15a (Fig. \ref{MOSFvar}, maroon triangles) shows a gentle increase in \Fvar\  toward lower energies.  The trend is similar for XMM15b (Fig. \ref{MOSFvar}, blue squares), but the changes between the soft and hard energy bands is more striking.   

Similar \Fvar\ spectra are seen in other Type I Seyferts, which can often be attributed to changes in photon index of a varying power law or changes in the ionization parameter of a reflector (e.g. Gallo et al. 2013, Fabian et al. 2013, Bonson et al. 2015).


\subsection{Principal Component Analysis}
\label{PCA}
A powerful tool used in many scientific fields for multi-variable statistical analysis is Principal Component Analysis (hereafter PCA). PCA performs an eigenvalue decomposition in which a data set is modelled by a number of linear relationships that minimize low or redundant information. These linear relationships are the principal components and are made up of an eigenvalue or coefficient (also called a `loading') along with its corresponding eigenvector. The principal components are not correlated by definition (see Feigelson \& Babu 1992 for review). In other words, PCA looks for a collection of related variables in a data set that explains most of the variance and clumps them together into a single principal component. It then repeats the process, finding the second most influential variables (giving the second principal component), and so on until a data set is reduced to a few principal components that best model variance in the data.

\begin{figure}
   \centering
   {\scalebox{0.35}{\includegraphics[trim= 0cm 0cm 0cm 0cm, clip=true]{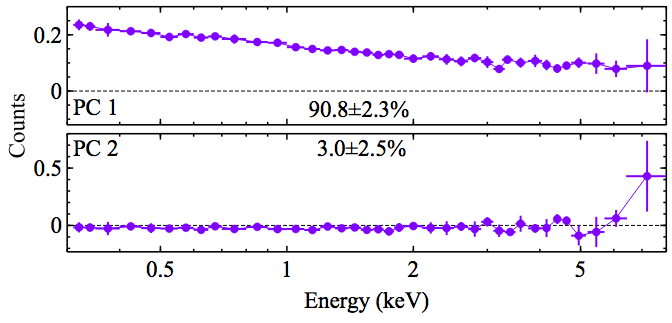}}}
   \caption{PCA normalized spectra using all the MOS data from 2003 and 2015. Data were binned by 10\ks.}
   \label{AvPCA2003to15}
\end{figure}

PCA is a powerful technique as it is model-independent and has the ability to reduce a data set with hundreds of parameters down to a much more manageable size while retaining information. In the case of AGN timing analysis, PCA can be applied to isolate the uncorrelated variable components in an X-ray spectrum and quantify their variability. A PCA spectrum can then be created in order to identify which of the original spectral component(s) may be responsible for the majority of the variability (e.g. Parker et al. 2014a, Parker et al. 2014b, 2015).

Using the $\textsc{pca\_public}$ code\footnote{\url{http://www-xray.ast.cam.ac.uk/~mlparker/}}, PCA was completed with data from all three \xmm observations: the 2003 and both 2015 epochs. Observations were divided in $10\ks$ intervals and a spectrum was created for each interval.  PCA was  conducted with data from: (i) the individual epochs, (ii) the 2015 epochs combined, and (iii) all three epochs combined. The PCA from each MOS camera at each epoch was consistent so the data were merged into a single spectrum.

Error bars were calculated by randomly perturbing the input spectra, recalculating the PCA, and finding the variance in the PCs themselves (Miller et al. 2007). The significance of individual components was assessed via log-eigenvalue (LEV) diagrams: the data were plotted by fractional variability as a function of eigenvector number. The data asymptotically approach a geometric series and components that fall above the trend outlined by the majority of points were considered significant (see Fig. 2 of Parker et al. 2014 as an example).

The PCA for the 2003 data was not informative given the short duration of the observation ($\sim 20\ks$).  The PCA of the individual 2015 data sets were found to be significant and comparable, therefore the data from XMM15a and XMM15b were combined. In the combined 2015 analysis the first three principal components were significant in the LEV diagram; however  PC 2 and PC 3 only showed variations above $\sim6\keV$ and may be attributed to high background.  PC1 accounted for 52.1$\pm$4.5 percent fractional variability and shows some sloping in the spectrum that decreases with increasing energy.

Finally, the high-flux state 2003 data are included with the low-flux 2015 data, and the PCA is calculated.  The results are comparable with that of the 2015 PCA  except that PC 1 is more significant accounting for 90.8$\pm$2.3 per cent of the spectral variability (Fig. \ref{AvPCA2003to15}). PC 2 is accounting for far less of the fractional variability in this case, $3.0\pm$2.5 per cent, with most of the variability at $> 5\keV$.  This could be intrinsic as some curvature is seen in the spectra (Fig. \ref{Spec}) or it could be due to increased background emission. 

Comparing Figure \ref{AvPCA2003to15} to simulations in Parker et al. (2015), it seems as though the primary source of both short-term (hours) and long-term (years) variability in \mrk493 is a change in normalization of some kind. The increased variability in the soft band is consistent with the results of the \Fvar\ analysis (Section \ref{Av_Fvar}) and in the difference spectrum that will be discussed in  Section~\ref{meanspec} (Fig. \ref{Spec}). 


\section{Preliminary examination of the mean spectra}
\label{meanspec}

We examine the average properties of the XMM03, XMM15a, and XMM15b spectra in a phenomenological manner.  
All spectral model fitting was performed using the X-ray spectral fitting package {\sc xspec v. 12.9.0}. Model parameters are reported in the rest frame of the AGN ($z=0.03$) and a cosmology of $H_{\rm{0}}$ = 70\kmpspMpc, $q_{\rm{0}}$ = 0, and $\Lambda_{\rm{0}}$ = 0.73 is assumed.  All fits include a Galactic column density of $N_{\rm{H}}$ = 2.11$\times$$10^{20}$\thinspace$\rm{cm^{-2}}$ as determined from the LAB Survey\protect\footnote{\url{http://heasarc.nasa.gov/cgi-bin/Tools/w3nh/w3nh.pl}} (Kalberla \et 2005) and modelled using the interstellar medium absorption model {\sc tbabs} and abundances from Wilms \et (2000).  Errors on model parameters correspond to a 90 per cent confidence level.

\begin{figure}
   \centering
   \advance\leftskip-0.5cm          
    {\scalebox{0.34}{\includegraphics[trim= 0.5cm 0.05cm 0.5cm 1cm, clip=true]{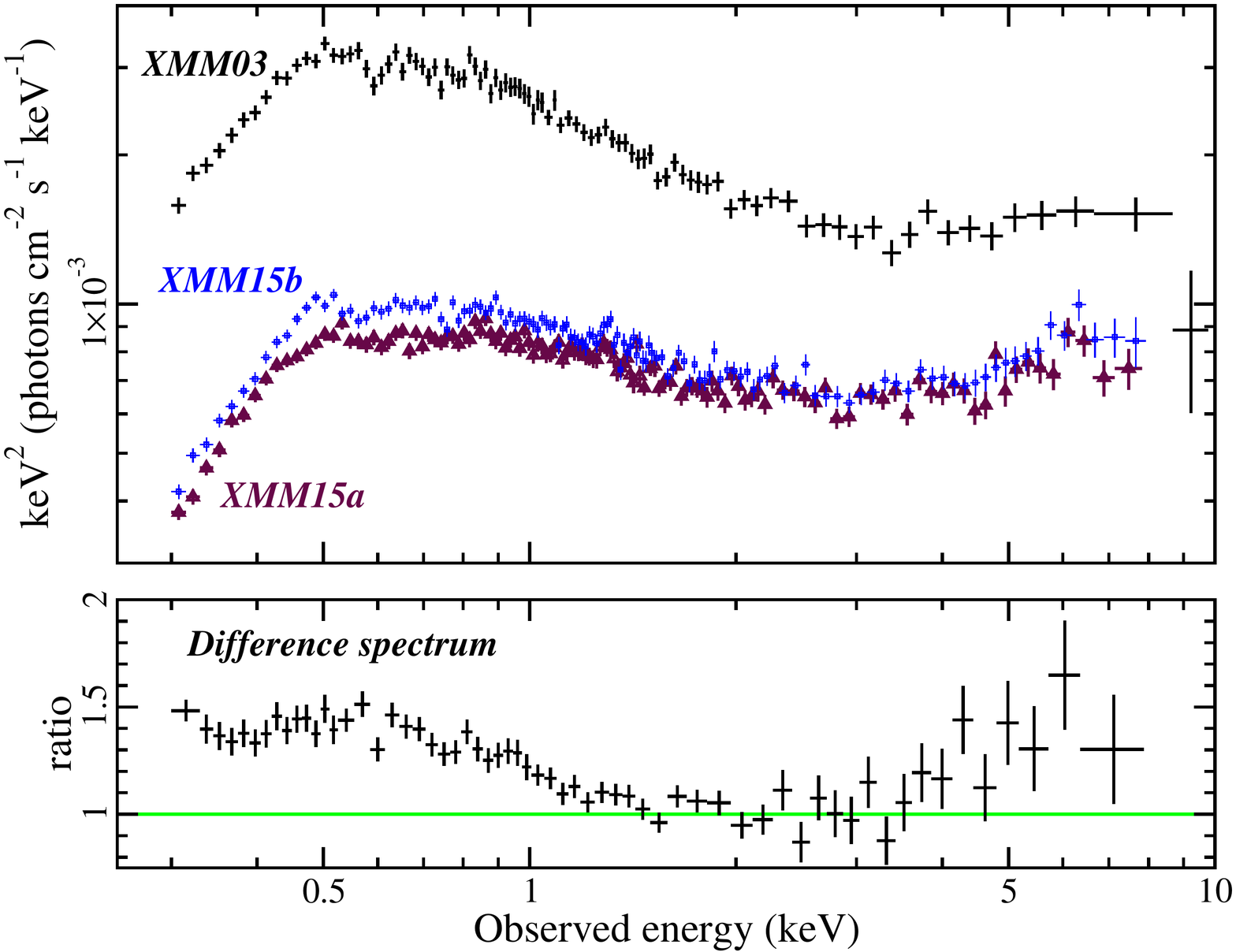}}}
    {\scalebox{0.34}{\includegraphics[angle=90, trim= 0.5cm 0.5cm 0.5cm 1cm, clip=true]{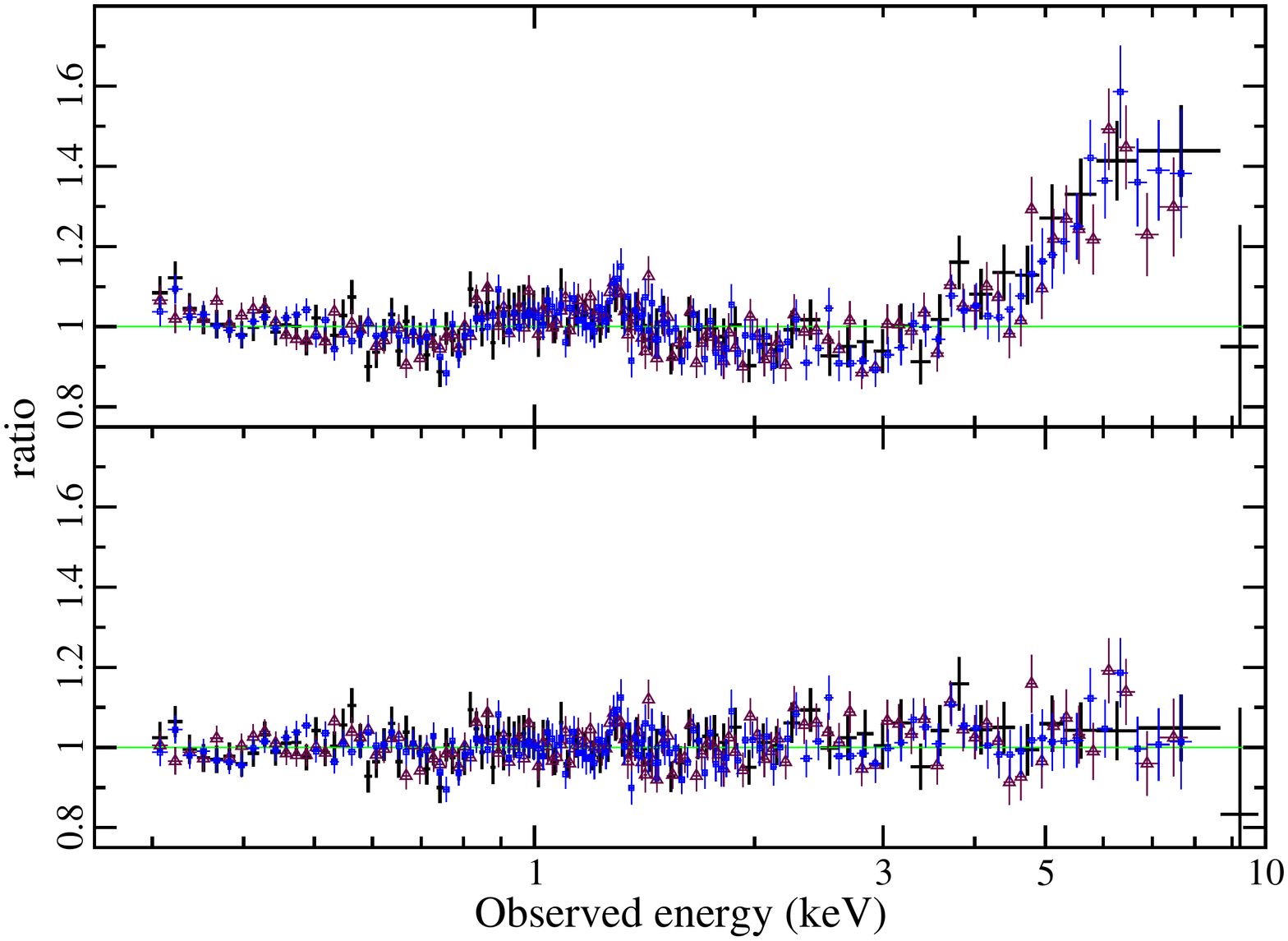}}}
   \caption{The unfolded spectra relative to a flat power law ($\Gamma=0$) are shown in the top panel.  The 2015 observations (maroon triangles and blue squares) catch Mrk~493 in a dimmer state than in 2003 (black crosses). The spectra are steep and a soft excess below $\sim1.5\keV$ is prevalent at all epochs.  The ratio of a difference spectrum (XMM03 -- XMM15b) fitted with an absorbed power law between $1.5-4.5\keV$ and extrapolated to higher and lower energies is shown in the second panel from the top.  The ratio from fitting the three spectra with a black body plus power law model shows excess residuals between $5-8\keV$ in all three epochs (second panel from bottom).  The lowest panel depicts the ratio from fitting the spectra with a broad Gaussian profile in addition to the black body plus power law continuum.}
   \label{Spec}
\end{figure}

In the top panel of Fig. \ref{Spec}, the three unfolded spectra relative to a flat power law ($\Gamma=0$) are shown to demonstrate changes between the epochs.  At all three epochs the spectrum is rather steep and a strong soft excess is clearly present.  The dimmer 2015 spectra show curvature with increasing energy peaking between $6-7\keV$.  Even though the 2015 spectra are obtained only 5 days apart, there are obvious changes in the soft excess during that time.   

A difference spectrum is created between the 2003 bright state and one of the 2015 spectra (XMM15b).  The difference spectrum is fitted with an absorbed power law ($\Gamma=2.7$) between $1.5-4.5\keV$ and extrapolated to lower and higher energies.  The ratio of this fit is shown in Fig. \ref{Spec} (second panel from top), clearly portraying changes in the soft excess and above $5\keV$ in the \feka\ band.

Each of the three spectra is fitted independently with a black body plus power law model.  The ratios of the fit are shown in Fig. \ref{Spec} (second panel from bottom).  Clear residuals exist in the \feka\ band that are improved when a broad Gaussian profile is added to each model (Fig. \ref{Spec}, lower panel).  Overall, the black body plus power law continuum with a broad Gaussian profile fits the three spectra well (\redchi\ $\sim 1.05$).  The parameters are typical of most NLS1s fitted in this manner.  The temperature of the black body component is $\approx 110\eV$.  The power law component is steep and changes significantly from $\Gamma \sim 2.68$ in 2003 to $\sim2.45$ in 2015.  The Gaussian profile is centred at $\sim 5\keV$ and is broad ($\sigma > 1\keV$) in all three spectra.  

Based on the best-fit model above the corresponding, unabsorbed fluxes in the $2-10\keV$ ($0.3-10\keV$) band are $3.57 \rm(15.2)$, $1.72 \rm(5.16)$ and $1.89 \rm(5.75)\times10^{-12}\ergpscmps$ for XMM03, XMM15a, and XMM15b, respectively. The $2-10\keV$ luminosity is approximately $3.6$ and $1.8\times 10^{42}\ergps$ in 2003 and 2015, respectively.

Notably, in 2015 the primary differences in the spectra lie in the soft excess.  While the flux of the broad Gaussian profile remains comparable in both 2015 spectra the soft excess (black body component) is about three-times brighter in XMM15b than it is in XMM15a.  The results from this simple examination of the spectra seems consistent with the the light curve analysis in Sect.~\ref{MIA} that show spectra variability is more significant at lower energies.

While differences are present between 2003 and 2015 (e.g. $\Gamma$ and normalization), from this point, the focus of the work will be on understanding the rapid spectral variability in the high quality 2015 data.

\begin{table}
\caption{The blurred reflection model for the simultaneous fit to the 2015 spectra. The merged-MOS data were fit between 0.3-8\keV. Parameters that are linked between epochs are shown with dots in column (4).  The reflection fraction ($R$) is the ratio of total fluxes (reflection/continuum) between $0.1-100\keV$. The reported unabsorbed flux is in the $2-10\keV$ band.}
\centering
\def\arraystretch{1.3}
	\begin{tabular}{cccc}                
	\hline
	(1) & (2) & (3) & (4) \\
 	Model   & Model   & XMM15a & XMM15b \\
	Component  &  Parameter   & &  \\
	\hline
 	Power  & $\Gamma$  & 2.14$\pm$0.03 & ... \\
	Law & model flux &  5.96$\times10^{-4}$ & ... \\
		 &   (\phpcmsps)  &  &  \\
	Blurred  & $q_{1}$ &  7.50$\pm$1.50 & $6.62^{+1.11}_{-2.88}$ \\
	 Reflector & $R_{\rm{in}}$ (\Rg)  & $1.25^{+0.34}_{-0.02}$ & ... \\
	 & $\theta$ (deg) & $53^{+6}_{-23}$ & ... \\
	 & $A_{\rm{Fe}}$ (solar)  & $0.53^{+0.09}_{-0.07}$ & ... \\
	 & $\Gamma$  &  $2.14$ & ... \\
	 & $\xi$ (\ergcmps)  & $1005^{+119}_{-242}$ & $800^{+228}_{-128}$ \\
	 & model flux  & 2.53$\times10^{-3}$ & 2.94$\times10^{-3}$ \\
	 & (\phpcmsps) &   &  \\
	 & \bf{$R$} ($F_{\rm{ref}}$ / $F_{\rm{pl}}$)  & 2.58$\pm$0.09 & 3.63$\pm$0.07 \\
	\\
	Flux  & $2-10\keV$ & $1.70\times10^{-12}$ & $1.83\times10^{-12}$ \\
	& ($\ergpscmps$)  & & \\
       Fit Statistic  &  \redchi\ /dof = 1.05 / 786 &   & \\
         \hline
\label{BR+DR_fit}
\end{tabular}
\end{table}

\section{The Blurred Reflection Scenario for 2015}
\label{Ref}
The analysis in the previous sections suggests the variability in 2015 is mainly driven by changes in normalization of a primary continuum component (e.g. Fig.~\ref{FvF} and \ref{AvPCA2003to15}), though the flux-flux plot in XMM15b indicate other factors are at work as well.  The X-ray spectra in the low-flux 2015 epochs is typical of most NLS1 and well described by a power law plus black body and a broad Gaussian profile at $\sim6\keV$ that resembles a relativistic iron line profile.  The predominant difference between the 2015 spectra is the strength of the soft excess.  Here, we consider if the blurred reflection model (e.g. Ross \& Fabian 2005) can describe the spectra and spectral changes seen in 2015.

\subsection{Fitting the 2015 average spectra}
\label{2015mean}

Relativistic blurred reflection is expected to be seen in objects in which there is an unobstructed view of the AGN central engine: namely, Type I sources. In essence, the primary, non-thermal X-rays from the corona are reprocessed in the accretion disk via a combination of processes including fluorescence, Compton scattering, and bremsstrahlung. This reflected emission is blurred due to the Doppler effects in the disk and relativistic effects close to the black hole. In addition, the extreme gravitational forces close to the supermassive black hole can enhance emission from the inner regions by light bending (e.g. Miniutti \& Fabian 2004). This boosted and smoothed reflection from the inner disk region forms the observed soft-excess emission below $\sim$\thinspace1\keV\ seen in most Type I sources.

\begin{figure}
   \centering
   \advance\leftskip-0.8cm              
   {\scalebox{0.34}{\includegraphics[angle=90,trim= 0.cm 1.5cm 1.cm 0cm, clip=true]{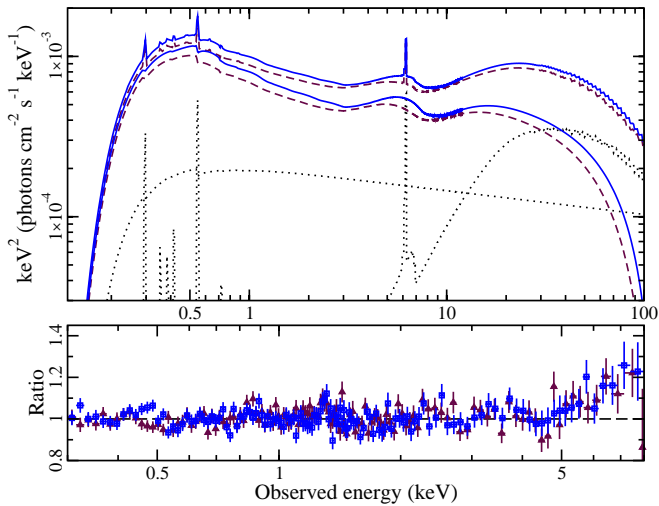}}} 
   \caption{Top: The best blurred reflection model, shown between $0.1-100\keV$, was simultaneously fit to the merged-MOS data of both 2015 observations. The model components corresponding to XMM15a are shown as maroon, dashed curves and those corresponding to XMM15b are shown as solid, blue curves.  The distant reflector and power law are shown as dotted, black curves.  These are non-variable and common to both epochs.  Bottom: The model residuals (data / model) between $0.3-8\keV$. Maroon triangles and blue squares correspond to XMM15a and XMM15b data, respectively.  Note the different energy scales used in the two panels.}
   \label{BestBR}
\end{figure}

Initially, spectra from XMM15a and XMM15b were examined separately to search for gross inconsistencies before fitting both spectra simultaneously.  The disk reflection model {\sc{reflionx}} (Ross \& Fabian 2005) was utilized.  The reflection spectrum was  modified for motions in the disc close to the black hole with the blurring kernel {\sc{kdblur2}}, which incorporates a broken power law for the emissivity profile. 
The following parameters were linked between the epochs: inner disk radius ($R_{\rm{in}}$), inclination angle ($\theta$), and iron abundance ($A_{\rm{Fe}}$).  The blurring parameters were also linked and the break radius ($R_{\rm{br}}$) and outer emissivity index ($q_{2}$), which were not well constrained when allowed to vary freely, were fixed to 10\Rg\ ($R_g= GM/c^2$) and 3, respectively.

In addition, both spectra seemed to require a narrow core at $\sim 6.4\keV$ that is modelled as distant, un-blurred reflection (e.g. from a torus). The components of the distant reflector are identical for both 2015 spectra.  To describe a neutral, distant reflector, such as that from a cold torus, the ionization parameter was fixed at $\xi$ = 1\ergcmps\ and the iron abundance was set to solar. The photon index of the power law source illuminating the distant reflector was fixed at $\Gamma$ = 1.9, the canonical value for AGN (e.g. Nandra \& Pounds 1994). There is no reason to assume that the radiation incident on the distant reflector will be the same as that from the primary component, but allowing $\Gamma$ to be linked to the power law continuum did not noticeably alter the fit.  The normalization was the only model component allowed to vary and it was linked between the epochs.  

Both spectra in 2015 are well-described with the single blurred reflector model (\chisq\ / dof = 825 / 784).  Slight positive residuals remained at $E>6\keV$, which may be due to the onset of significant background emission or true spectral variability.  A variable component above $6\keV$ is seen in the PCA (Fig.~\ref{AvPCA2003to15}).  We also note the spectrum above $6\keV$ is better fit when we consider time-resolved spectroscopy in Section~\ref{TRS_BR}, suggesting the difficulty fitting the average spectrum may be from rapid spectral variability. 

If the primary components ($\Gamma$ and normalization) were linked between XMM15a and XMM15b, the fit quality was comparable (\chisq\ / dof = 826 / 786) and the residuals are unchanged. Therefore, in the final model the power law component is linked between the epochs (Fig. \ref{BestBR}; Table \ref{BR+DR_fit}).

The parameters for the best-fit blurred reflector model are listed in Table \ref{BR+DR_fit}. Based on the fits to the two average spectra, \mrk493\ is reflection dominated in 2015 when it is at a relatively low-flux interval.  The  difference between XMM15a and XMM15b can be explained primarily with changes in the flux and ionization parameter of the blurred reflector.  We examine this in further detail in the following sections.

The model-predicted $14-195\keV$ flux is approximately $2.4\times10^{-12}\ergpscmps$, which is consistent with the null-detection in the 105-month \swift-BAT survey (Oh \et 2018).  


\subsection{Time-Resolved Spectroscopy}
\label{TRS_BR}

\begin{figure*}
   \centering
   {\scalebox{0.36}{\includegraphics[trim= 1cm 1cm 3cm 1cm, clip=true]{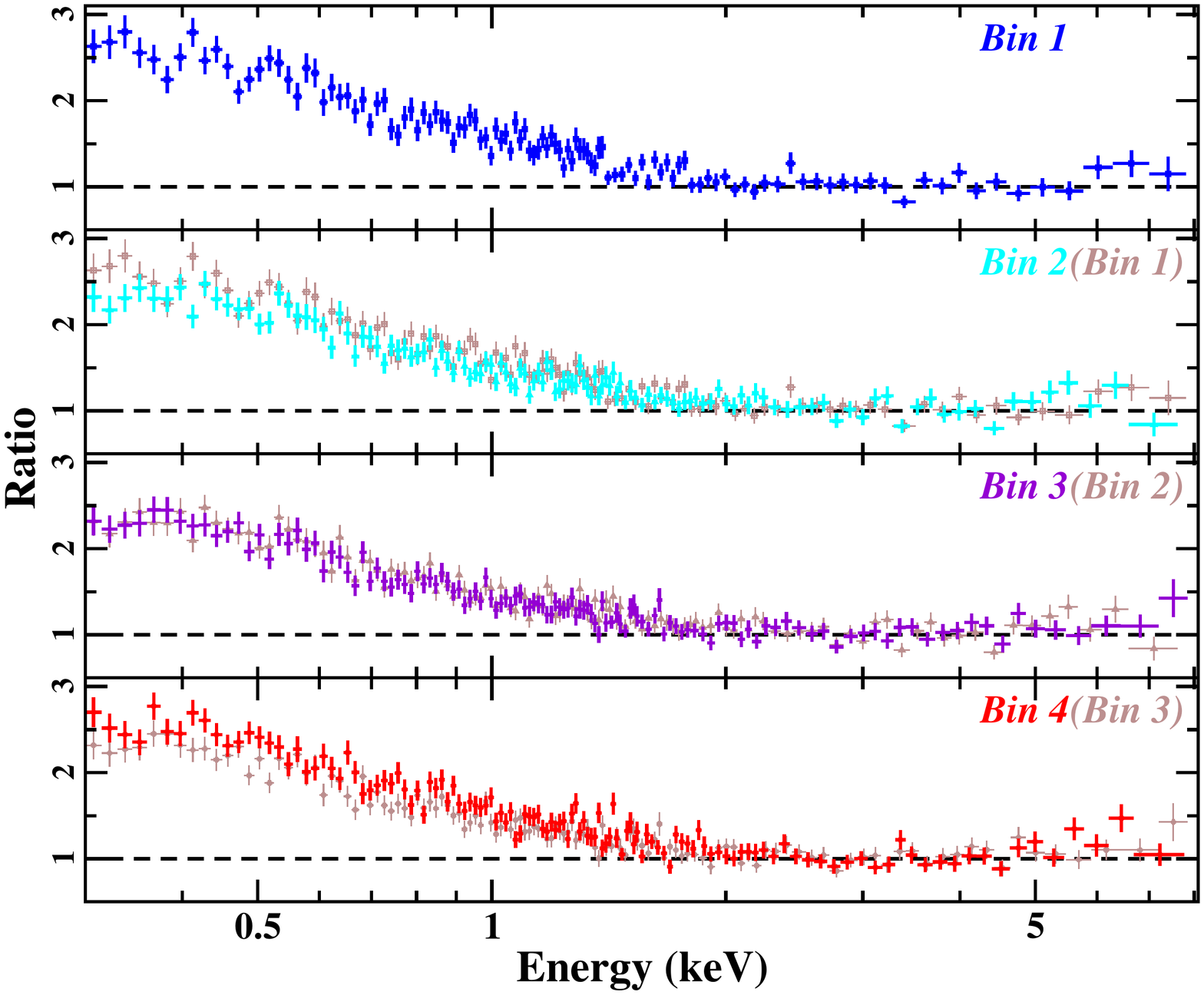}}}
   {\scalebox{0.36}{\includegraphics[trim= 1cm 1cm 3cm 1cm, clip=true]{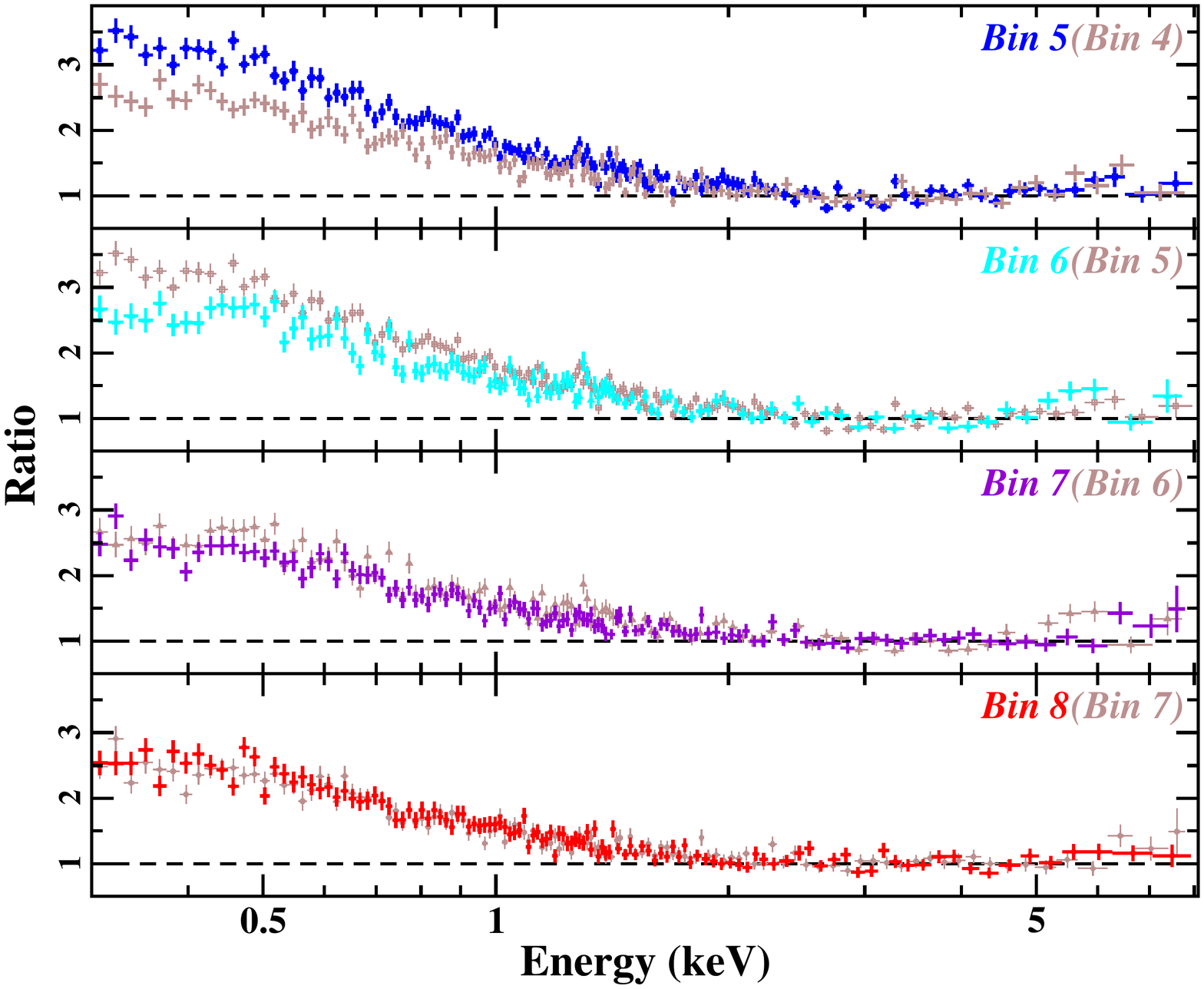}}}
   \caption{A power law model is fit between $2-8\keV$ to all time-resolved spectra while keeping the photon index linked and the normalisation free to vary.  The model is extrapolated to $0.3\keV$ and the residuals (ratios) are shown for the time bins in XMM15a (left panel) and XMM15b (right panel).  The residuals in each bin are compared to those in the preceding time segment (faded brown data)  to highlight changes from one spectrum to the next.  }
   \label{101TRS}
\end{figure*}

Time-resolved spectra are created for the consecutive $25\ks$ segments that are shown in Fig.~\ref{TRL}.  There are four spectra created for XMM15a (numbered $1-4$) and four spectra for XMM15b (numbered $5-8$) (see Fig.~\ref{TRL} and \ref{101TRS}).

An absorbed power law is fit to each time-resolved spectrum between $2 - 8\keV$.  The photon index is linked between the time bins at each epoch, while the normalization is allowed to vary.  The fit is then extrapolated to $0.3\keV$.  The residuals from this fit are displayed in Fig. \ref{101TRS}, showing the spectral changes as a progression from one bin to the subsequent bin.
In addition to changes in power law flux, Fig. \ref{101TRS} demonstrates there are also difference in the soft-excess (e.g. Bin 1 - 2 and Bin 4 - 5) and in the \feka\ band (e.g. Bin 5 - 6 and Bin 6 - 7).  Interestingly, there is some indication that changes in the soft-excess may precipitate changes in the \feka\ band. For example, increased soft-excess emission in Bin 5 is followed by enhance emission between $5-6\keV$ in Bin 6.

The blurred reflection model is used to fit the eight time-resolved spectra.  Considering the parameters that changed in fitting the 2015 average spectra (Sect.~\ref{2015mean}), the power law normalization, inner emissivity index,  disk ionization, and reflector normalization are permitted to vary between time bins. Allowing the photon index to be free also improves the fit, though the value of the parameter did not change significantly from one spectrum to the next.  The model reasonable fit all eight time-resolved spectra (\chisq/dof = 1945 / 1899). 

A Markov Chain Monte Carlo (MCMC) method was used to estimate the uncertainties for the time-resolved spectral model via an open-source algorithm developed by Foreman-Mackey et al. (2013) utilizing the original  Goodman \& Weare (2010) method. The \textsc{xspec}-friendly program, \textsc{xspec\_emcee}, was developed by Jeremy Sanders and is publicly available.\footnote{\url{https://github.com/jeremysanders/xspec_emcee}} The MCMC fitting was run with 76 walkers and 10,000 iterations. The best-fit parameter values were used as the peak of the MCMC probability distributions and the errors were the widths of those distributions. The first 1000 steps were burned to ensure truly random initial conditions. Only the photon index, power law normalization, blurred reflection ionization, and the inner emissivity index were allowed to vary. The inclination angle and iron abundance were linked between all the spectra, but allowed to vary. The flux of the power law and that of the blurred reflector were calculated between 0.1 -- 100\keV\ to track reflection fraction ($R$) as well.

\begin{figure*}
   \centering
   {\scalebox{0.36}{\includegraphics[trim= 1cm 1cm 3cm 1cm, clip=true]{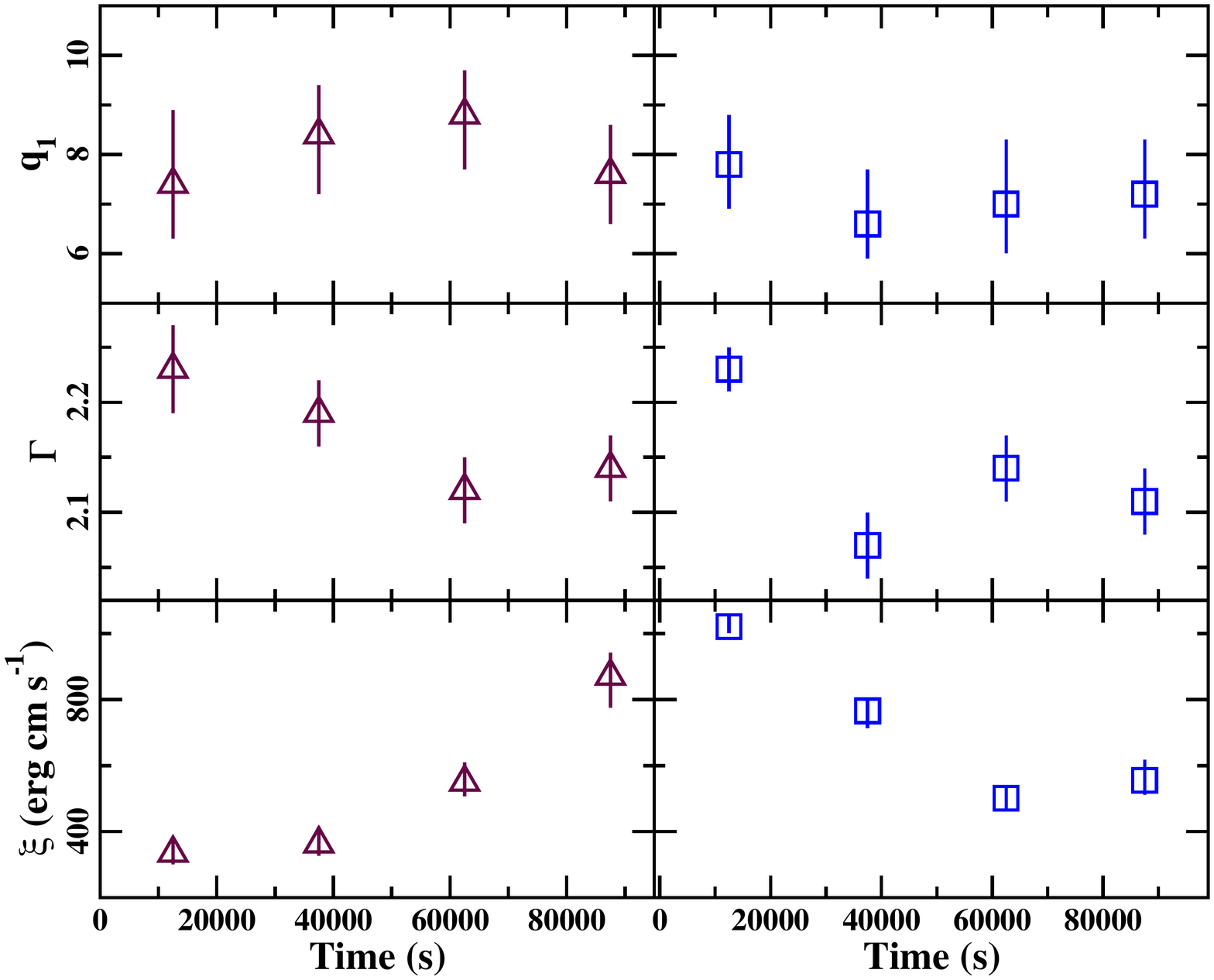}}}
   {\scalebox{0.36}{\includegraphics[trim= 1cm 1cm 3cm 1cm, clip=true]{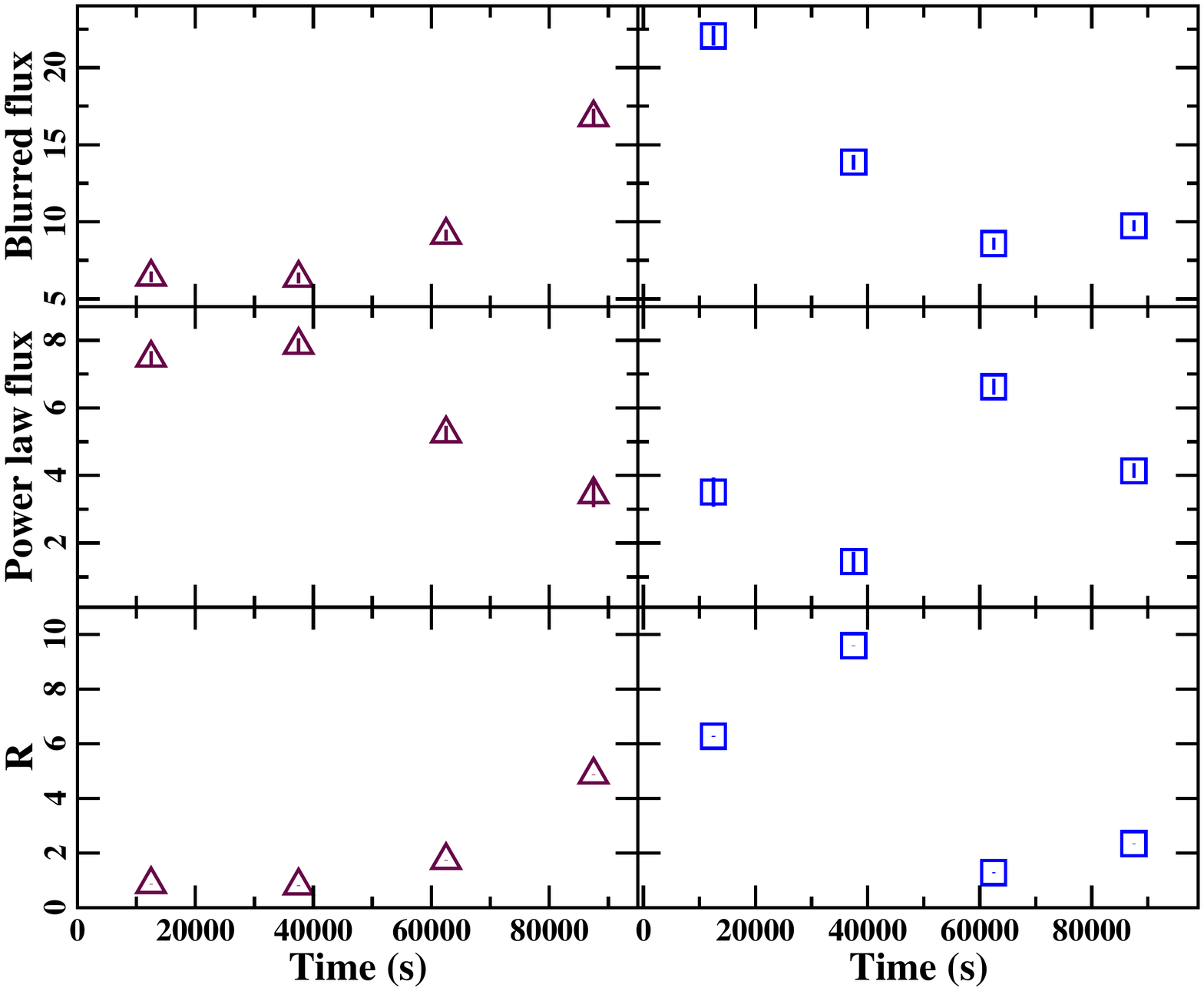}}}
   \caption{Trends extracted from the simultaneous fitting of time-resolved spectra with the blurred reflection model. 
   Data from XMM15a (maroon triangles) and XMM15b (blue squares) are binned in approximately $25\ks$ bins as shown in Fig. \ref{TRL}.  Power law and blurred reflector fluxes are given in units of
   $10^{-4}$ and $10^{-3}$ ph~cm$^{-2}$~s$^{-1}$, respectively.}
   \label{SeriesA}
\end{figure*}

A time series was created for each fit parameter to track changes from one bin to the next (Fig.~\ref{SeriesA}).   
We see that while the inner emissivity index remains relatively constant within uncertainties, the other three parameters show significant variability across the 25\ks\ time bins.   The power law photon index does vary, but by only about $\pm5$ per cent.
Disk ionization more than doubles between the start of XMM15a and the start of XMM15b, before dropping again to nearly its starting values. Power law normalization shows an opposite trend to that of disk ionization: dropping from its starting value to a minimal value in Bin 2, before climbing once again. 

\begin{figure}
   \centering
   \advance\leftskip-0.5cm          
       {\scalebox{0.35}{\includegraphics[trim= 1cm 1cm 3cm 3cm, clip=true]{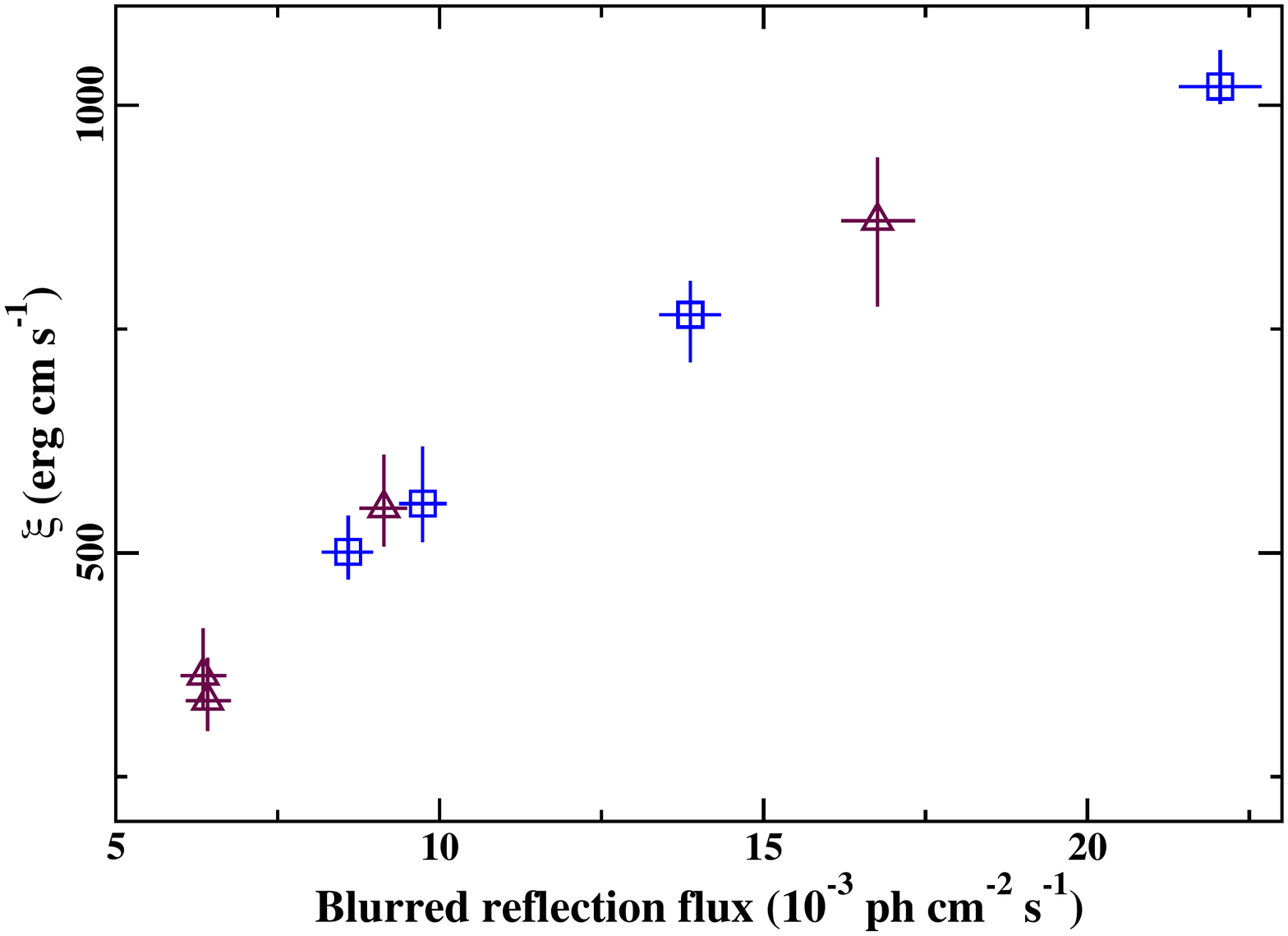}}}
   {\scalebox{0.35}{\includegraphics[trim= 1cm 1cm 3cm 2cm, clip=true]{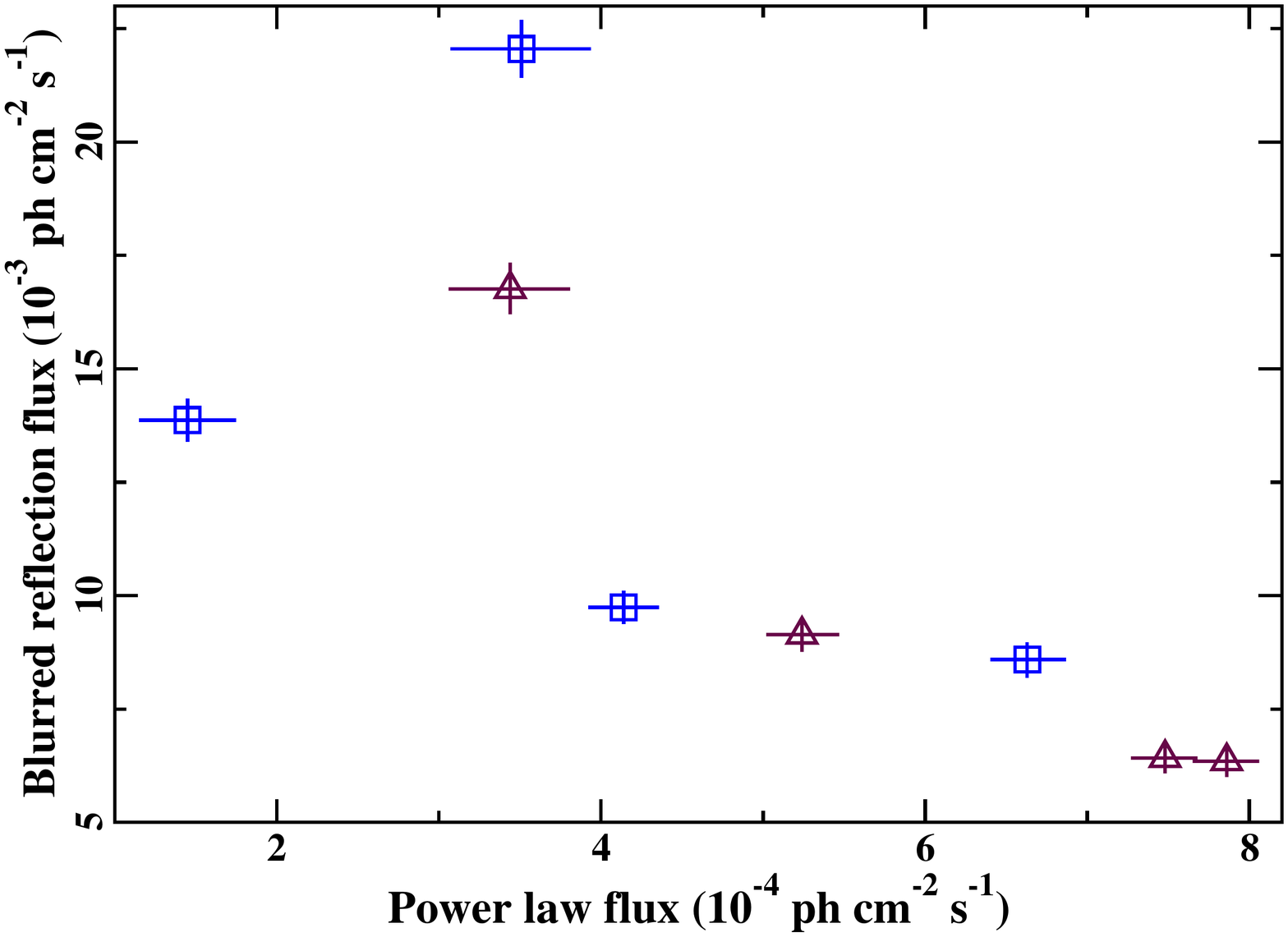}}}
   \caption{Comparison between measured parameters in the time-resolved spectral analysis.  The disk ionization parameter is compared to the flux from the blurred reflector in the top panel.
   The blurred reflector flux and power law flux are compared in the lower panel.  Data from XMM15a and XMM15b are shown as maroon triangles and blue squares, respectively.}
   \label{mcmc}
\end{figure}

The most variable parameters in Fig.~\ref{SeriesA} are compared to each other to search for correlations (Fig.~\ref{mcmc}).  There is a clear inverse trend between the flux of the power law component and the flux of the blurred reflector (lower panel of Fig.~\ref{mcmc}) and a tight correlation between the ionization parameter and the blurred reflector flux (upper panel of Fig.~\ref{mcmc}).  The trends are consistent with the light bending interpretation (e.g. Miniutti \& Fabian 2004).  If photons from power law component are directed toward the disk due to the curve space-time around the black hole, the direct power law flux reaching the observer drops as more photons strike the inner disk, increasing the flux from the reflection component.  Since more photons are striking the inner disk, the ionisation parameter will consequently increase as well.  The trend is also apparent in the reflection fraction ($R$), which is highest during moments of low power law flux.  On average, $R>1$ indicating Mrk~493 is reflection dominated.

The described light bending scenario could occur in a simple lamp-post geometry where the source (power law emitter) height above the disk is changing.  We considered this possibility by modelling the spectra of Bin 1 and Bin 5, which have diverse blurred reflection parameters, with the {\sc xspec} model {\sc relxilllp} (Garcia \et 2014).  The  source height is a free parameter in the  {\sc relxilllp} model.  Unfortunately, the model could not significantly distinguish different source heights in Bin 1 and 5.  While the resulting values are consistent with the scenario above (i.e. lower source height in Bin 5) the measurements are comparable within uncertainties.

\subsection{Time-Resolved \Fvar\ Modelling}
\label{TRFvar_Mod}

\begin{figure*}
   \centering
   {\scalebox{0.25}{\includegraphics[trim= 0.4cm 0.4cm 0.4cm 1cm, clip=true]{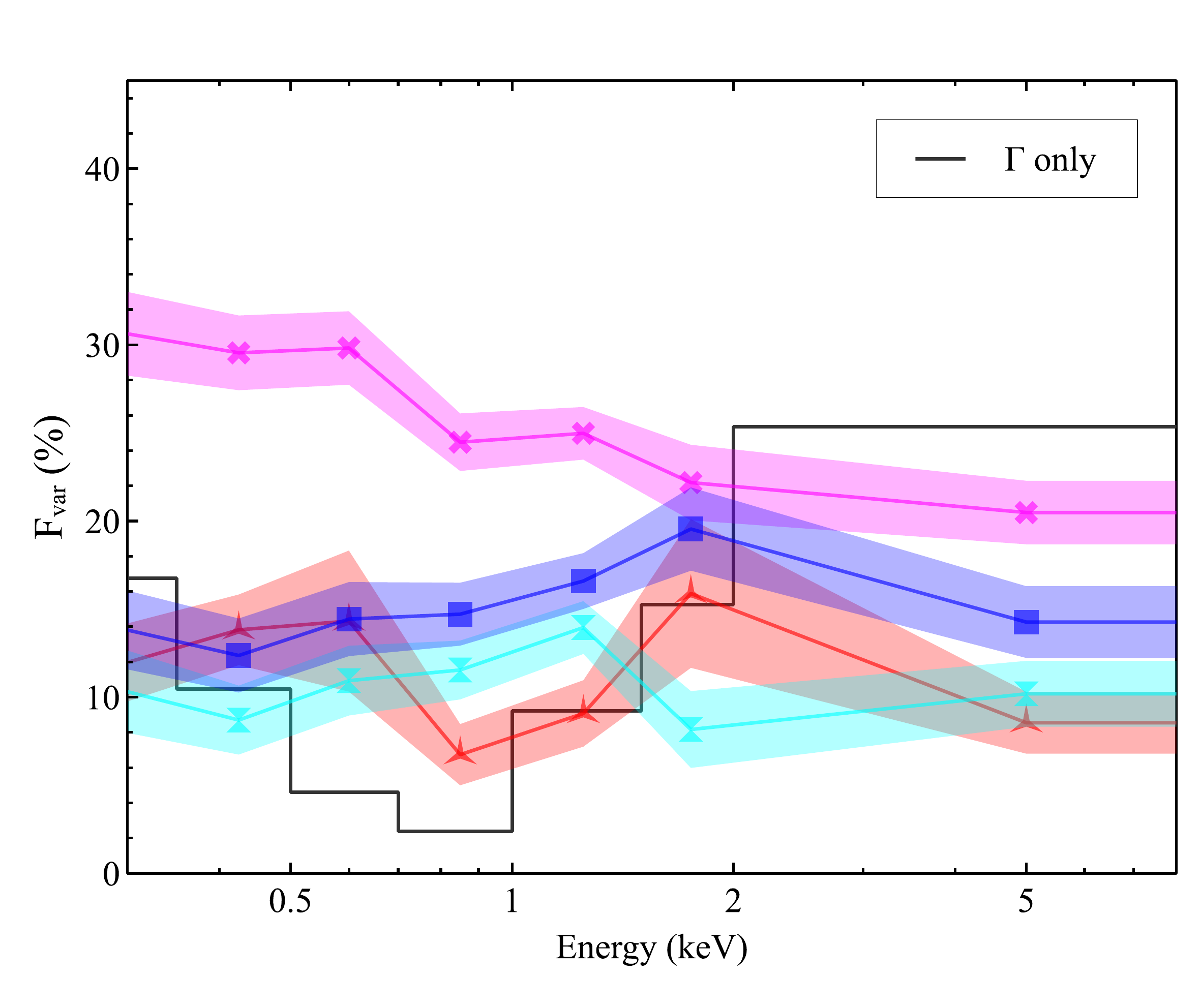}}}  
      {\scalebox{0.25}{\includegraphics[trim= 0.4cm 0.4cm 0.4cm 1cm, clip=true]{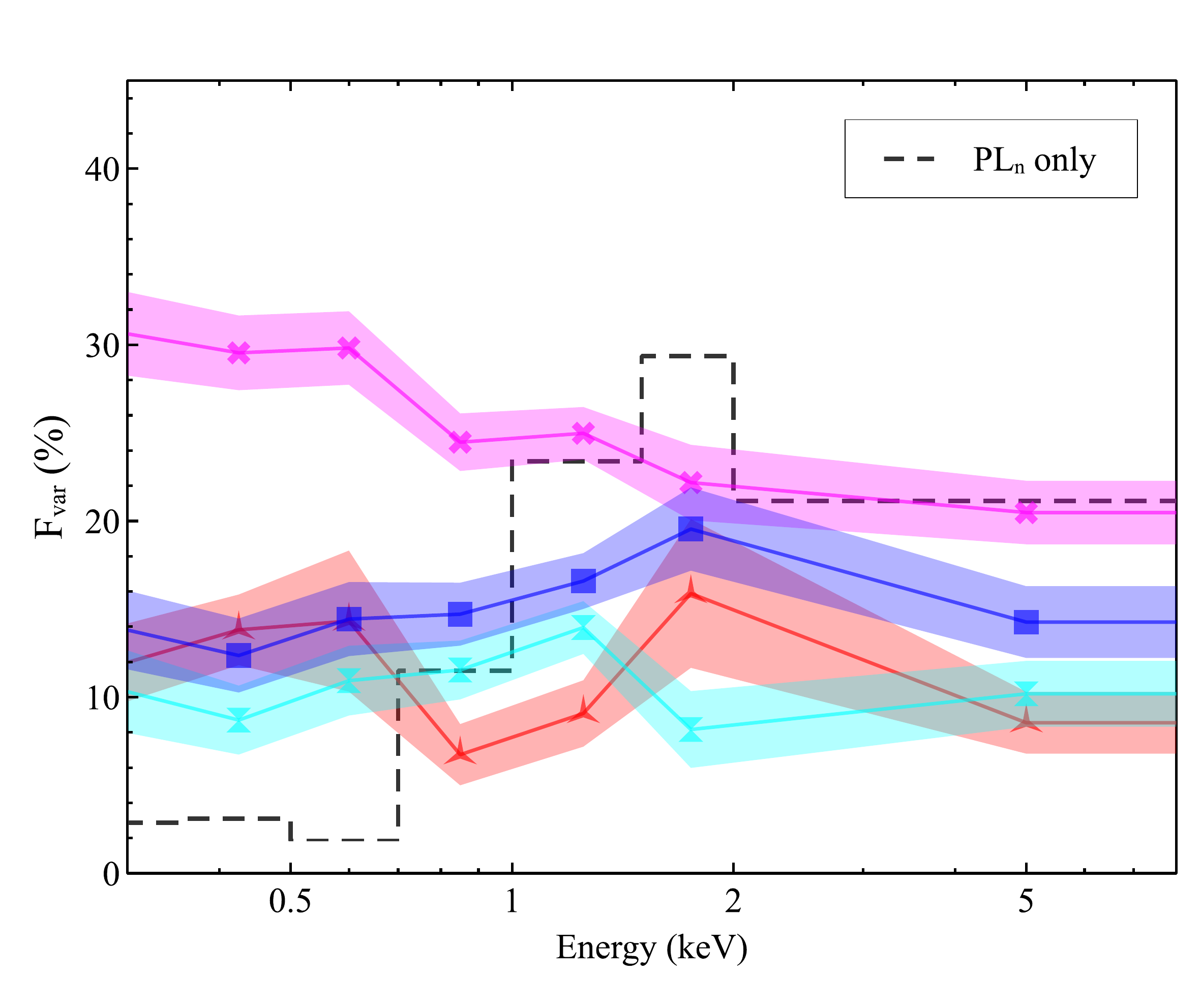}}}   
         {\scalebox{0.25}{\includegraphics[trim= 0.4cm 0.4cm 0.4cm 1cm, clip=true]{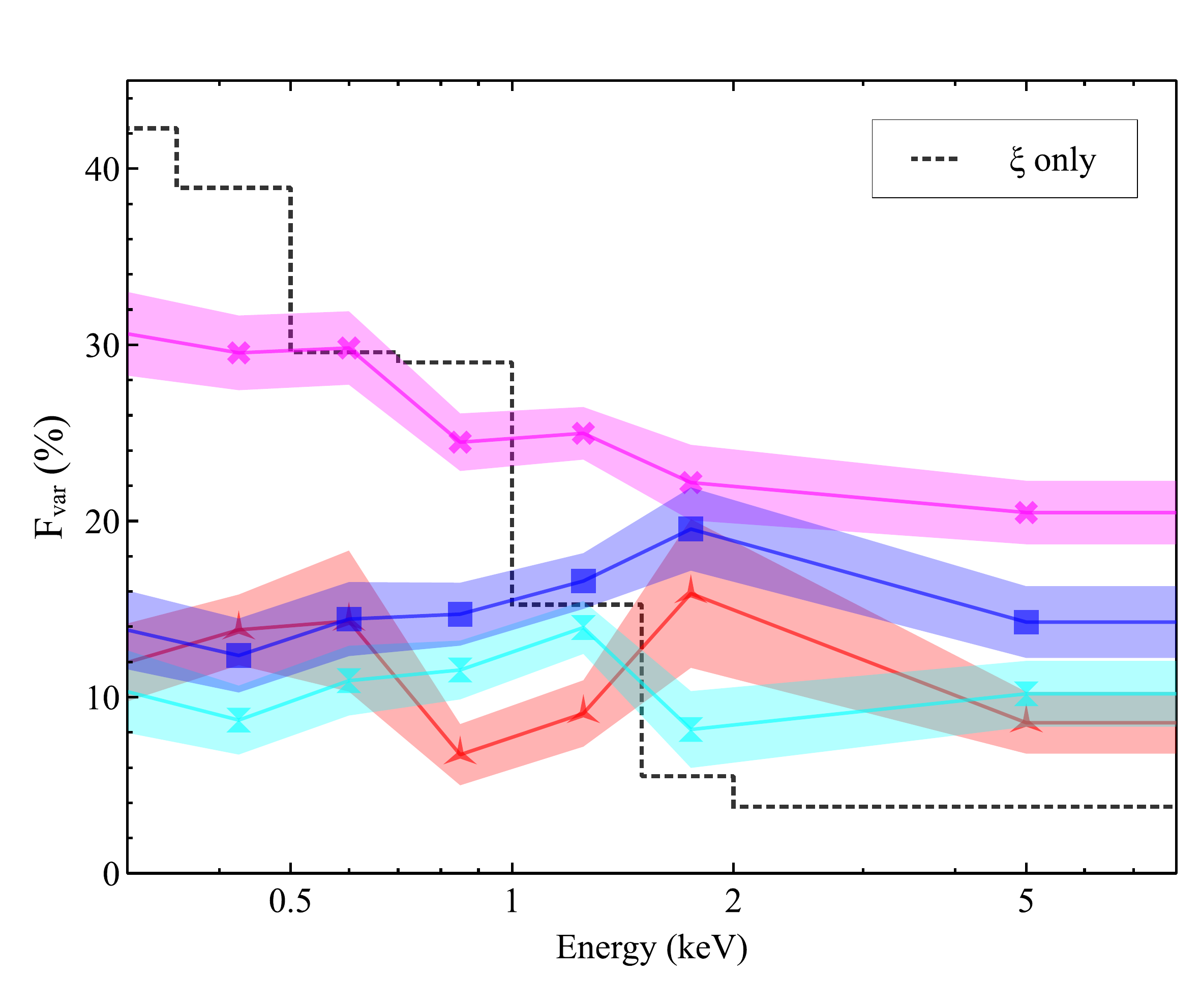}}}    
   {\scalebox{0.25}{\includegraphics[trim= 0.4cm 0.4cm 0.4cm 1cm, clip=true]{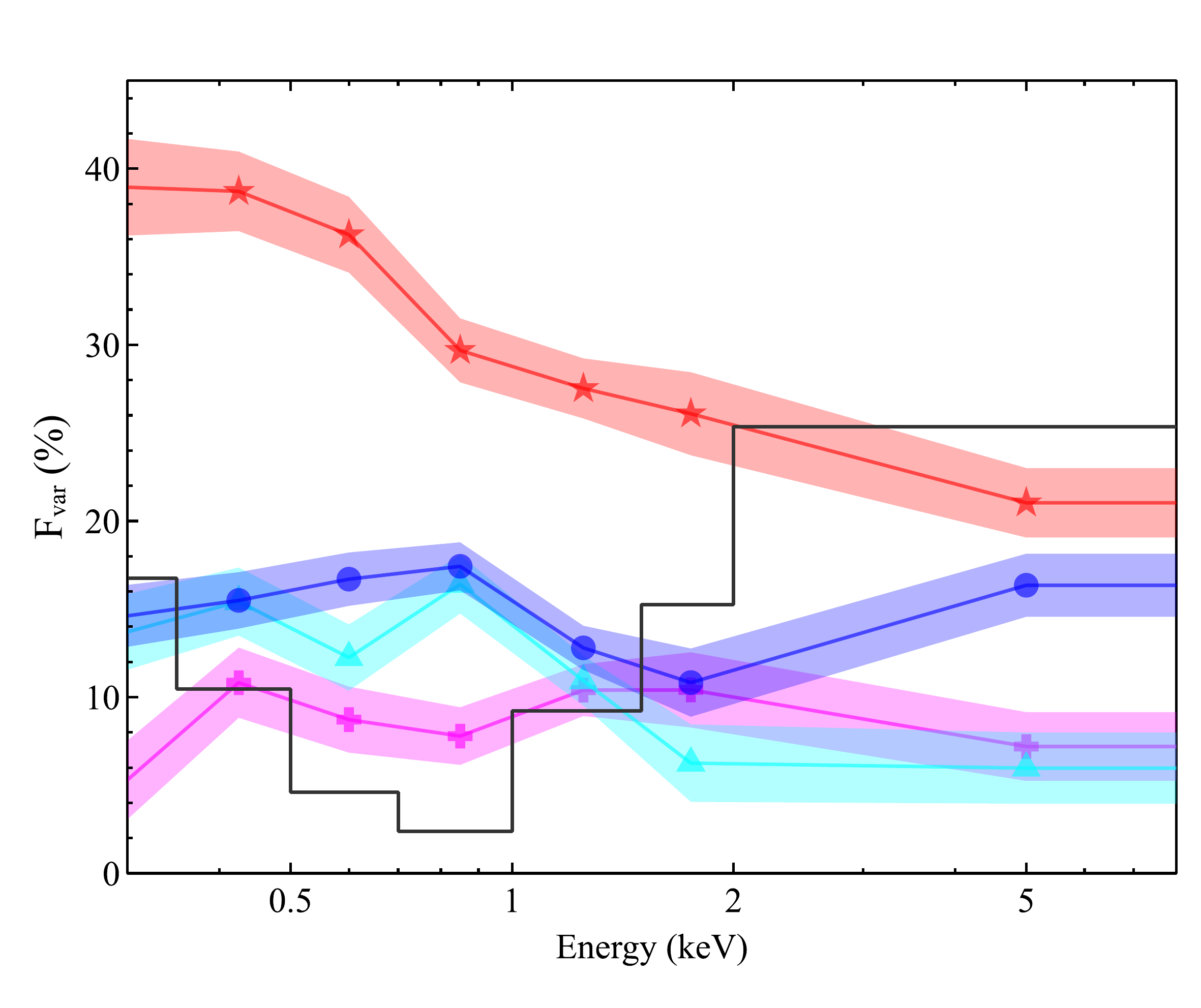}}}    
   {\scalebox{0.25}{\includegraphics[trim= 0.4cm 0.4cm 0.4cm 1cm, clip=true]{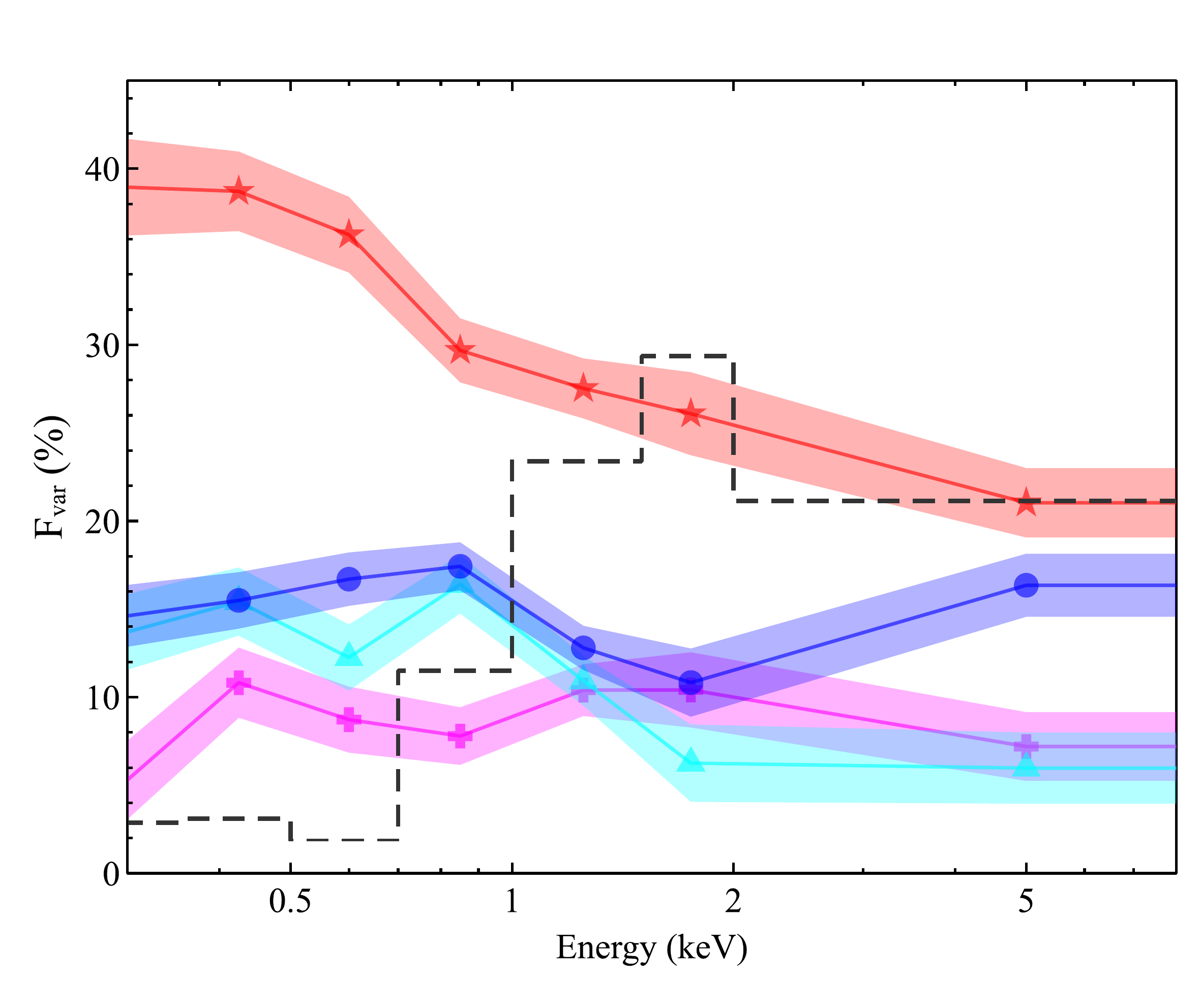}}}  
   {\scalebox{0.25}{\includegraphics[trim= 0.4cm 0.4cm 0.4cm 1cm, clip=true]{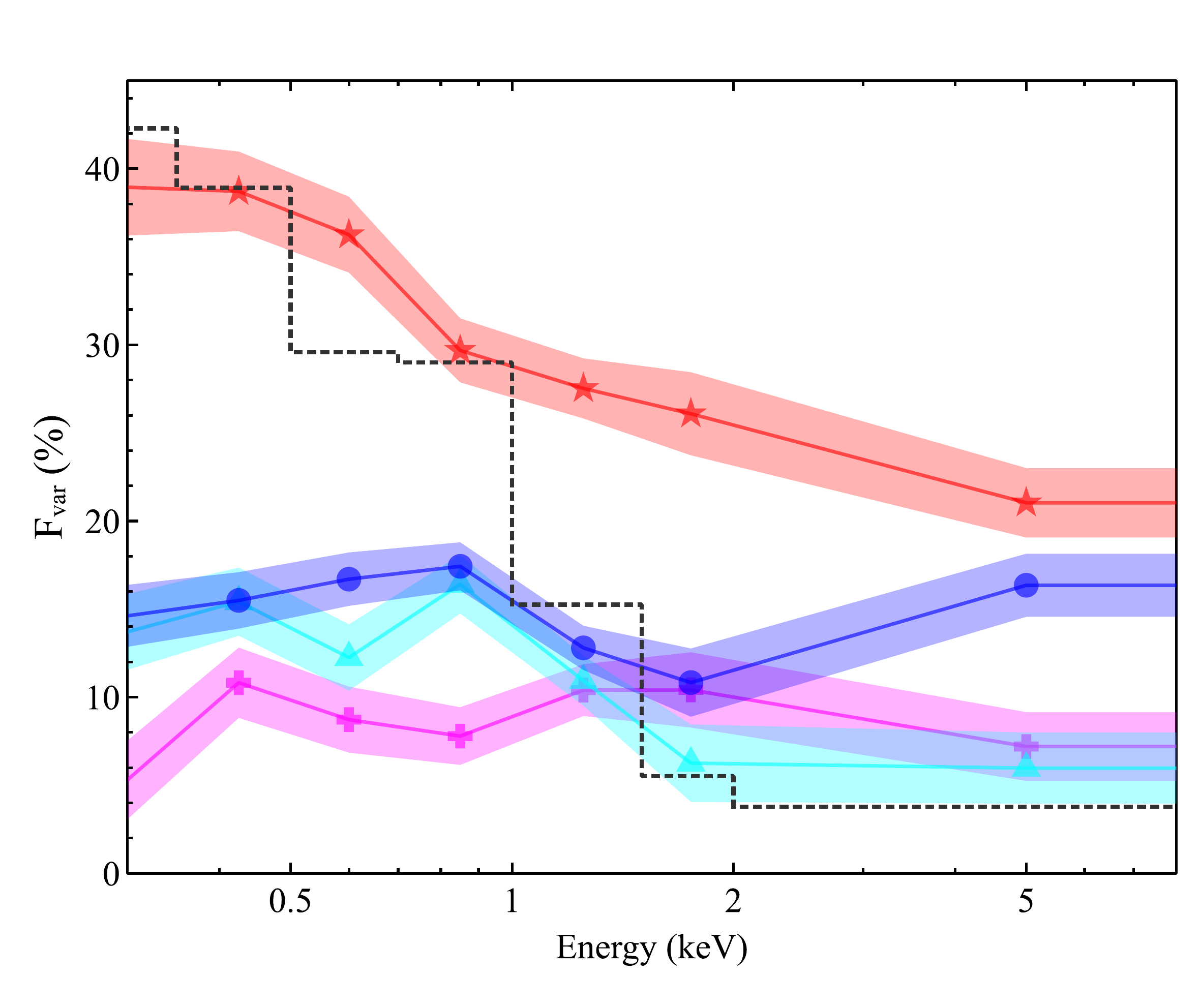}}}    
   \caption{Time-resolved \Fvar\ spectra for each time segment shown in  Fig. \ref{TRL}.  Data of XMM15a and XMM15b are shown in the top and lower panels, respectively.  Symbols and colours correspond to the segments as shown in Fig. \ref{TRL}.  The most variable segments correspond to Bin 3 (purple crosses) during XMM15a and Bin 8 (red stars) during XMM15b.  Theoretical \Fvar\ spectra are overplotted as black curves in each panel.  The inset in the top panels identifies the parameters that are assumed to vary to created the simulated \Fvar\ for that column.  From left-to-right, the variable parameters are the photon index ($\Gamma$, solid), power law normalization ($PL_{\rm{n}}$, dashed), and disk ionization ($\xi$, dotted).  A simple scenario in which only one model parameter changes at a time does not reproduce the spectral variability in \mrk493.
 }
   \label{TRFvar_BRMod}
\end{figure*}

Fractional variability spectra were created for each of the eight time segments to investigate how the \Fvar\  changes over the 2015 observations.    These time-resolved \Fvar\ spectra are shown in Fig. \ref{TRFvar_BRMod} using the same symbols and colours as defined in Fig. \ref{TRL} to identify data in different time segments.  The combination of the time-resolved spectra appear generally consistent with the average 
\Fvar\ spectra in Fig~\ref{MOSFvar}.  

The time-resolved \Fvar\ spectra are approximately flat during all time segments, but the most significant spectral variability and largest amplitude variations are seen in Bin 3 and Bin 8.  The segments  correspond to periods when \mrk493\ was at its lowest average flux.  During these more active segments, the amplitude of the variations are enhanced at all energies, but the majority of the variability is in the soft excess below $\sim$\thinspace1\keV.

To investigate possible scenarios that could produce the different time-resolved \Fvar\ spectra, we simulate one hundred spectra based on the best-fit blurred reflection model (Sect.~\ref{2015mean}) allowing only one parameter to vary randomly.  The parameters tested were the photon index ($\Gamma$), power law normalization ($PL_{\rm{n}}$), and the disk ionization parameter ($\xi$).  From these simulated spectra, theoretical \Fvar\ spectra were calculated and overplotted on the time-resolved \Fvar\ spectra of \mrk493\ (Fig. \ref{TRFvar_BRMod}).

It is not possible to reproduce the flat \Fvar\ spectra seen during most of the time segments by allowing only one parameter to vary at a time.  Most likely, multiple parameters need to vary together in some manner to reproduce the constant variations across the entire energy band.  For example, Fig.~\ref{SeriesA} shows how the blurred reflection and power law flux (i.e. normalization) are anti-correlated during the observation.

The enhanced variability at lower energies is most likely achieved by changing the ionization parameter.  Varying the photon index can not reproduce the large amplitude changes at low energies since the soft component (i.e. blurred reflection in our model) would  dampen the variations.  Again, the mostly likely scenario is that multiple components need to change simultaneously to accurate recreate the \Fvar\ spectra.

\section{Discussion} 
\label{discuss}

This is the first in-depth examination of the NLS1 galaxy, \mrk493, with \xmm.  Two, approximately $100\ks$ observations separated by $\sim5$~days, show the source to have a typical NLS1 X-ray spectrum with a strong soft excess and broad excess emission in the \feka\ band.  The average spectra are well-described by a blurred reflection model that indicates the AGN is reflection dominated, like many NLS1s (e.g. Fabian \et 2009, 2013; Gallo \et 2004).  

The AGN appears to be in a low flux interval, which would be consistent with the reflection dominated scenario (e.g. Gallo 2006).  For example, the 2015 data of \mrk493\ were  compared to a 2003 snap-shot observation when the AGN was more than twice as bright (Fig.~\ref{Spec}).  The PCA utilizing 2003 and 2015 data (Fig.~\ref{AvPCA2003to15}) is compared to models generated by Parker \et (2015) and suggest the primary changes are in the normalization and pivot of the power law component.  

Time-resolved spectroscopy, PCA, flux-flux plots, and \Fvar\ modelling indicate the short-term variability during the 2015 \xmm\ observations can not be described by variability in a single component.  Changes in the normalization of the power law and blurred reflector are necessary, but strong changes in the soft excess emission suggests the ionization parameter of the blurred reflector fluctuates as well. 

Interestingly, changes are also seen in the \feka\ band.  For example, in Bin 6 of the time-resolved spectroscopy (Fig.~\ref{101TRS}), enhanced emission is seen between $5-6\keV$ that succeeds changes in the soft excess in Bin 5.  These changes in the \feka\ emission could originate along with overall changes in the continuum, but could also arise from spots or annuli on the disk that momentarily brighten or infalling material (e.g. Bonson 2017, Yaqoob et al. 2003, Giustini et al. 2017).

The time-resolved spectroscopy shows the varying model parameters are the blurred reflector and power law flux, and the ionization parameter.  Comparing the parameters, we find a strong correlation between the ionization parameter and the flux of the blurred reflector as well as a negative trend between the flux of the reflector and power law (Fig.~\ref{mcmc}).  These trends seem consistent with the light bending scenario (e.g. Miniutti \& Fabian 2004).  If we consider a standard lamp-post geometry where the primary source is an isotropic emitter of roughly constant luminosity situated above the black hole, but allowed to vary in height, then when it is in the strong gravity environment close to the black hole, most of the power law photons will be directed toward the inner accretion disk.  The primary (power law) source will appear dimmer to the distant observed while the reflected emission will appear relatively enhanced.  As more continuum photons are striking the inner accretion disk, the ionization parameter of the reflector will increase producing the correlations we observe in Fig.~\ref{mcmc}.

The described scenario  predicts changes in the inner emissivity index ($q_1$) of the disk that are not clearly observed in the time-resolved spectra (Fig.~\ref{mcmc}).  However, this could be a data quality issue.  The parameter $q_1$ is  poorly constrained with the current data.  Values in the time-resolved analysis are uncertain at about $\pm20$ per cent and best-fit values range from $q_1 \sim6-9$.  In addition, the break radius was not well constrained when it was free to vary so it was fixed at $10$\Rg.  The break radius and emissivity index are coupled in the lamp-post scenario and the break radius will become smaller when $q_1$ increases.

Though the primary source is treated as a point in this lamp-post scenario, it could equivalently be described as a compact spherical corona or the base of a jet.  Both structures could be rather dynamic generating the power law variability (e.g.  Wilkins \et 2014, 2015; Wilkins \& Gallo 2015) driving the changes in the blurred reflector.  Distinguishing these origins is possible (Wilkins \& Fabian 2012; Gonzalez \et 2017) and should be attempted with higher quality data in the $2-10\keV$ band.

\section{Conclusions} 
\label{conclude}
We examined the characteristics of the X-ray continuum of the NLS1 galaxy, \mrk493, by considering both the long- and short-term variability. Data from \xmm merged-MOS instruments were primarily utilized due to high background  in the pn, however the pn data were processed alongside all investigations to confirm results. 

The 2015, low-flux spectra of \mrk 493\ are well-fit with a blurred reflection model that is reflection dominated.   However, the variability within the 2015 observation is complex and can not be described by fluctuation in a single component.  In addition to changes in brightness of the power law and reflector, variations in the ionisation parameter of the blurred reflector are needed to fit the soft excess.

The behaviour displayed by \mrk493\ in 2015 suggests the X-ray source is compact and that the NLS1 will be a useful source for studying the inner accretion disk around supermassive black holes.

\section*{Acknowledgments}
KB would like to thank A. G. Gonzalez and H. Ehler for thoughtful brainstorming sessions.  Thanks to Ciro Pinto for discussion and to the referee for comments that improved the paper.
The \xmm\ project is an ESA Science Mission with instruments and contributions directly funded by ESA Member States and the USA (NASA).
ACF acknowledges ERC Advanced grant 340442.

\bibliographystyle{mn2e}





\end{document}